%% file: main.tex
\documentclass[sigconf,9pt]{acmart}

\input{preamble/preamble}

\input{preamble/macros}

\copyrightyear{2026}
\acmYear{2026}
\setcopyright{cc}
\setcctype{by}
\acmConference[IMC '26]{Proceedings of the 2026 ACM Internet Measurement Conference}{October 12--16, 2026}{Karlsruhe, Germany}
\acmBooktitle{Proceedings of the 2026 ACM Internet Measurement Conference (IMC '26), October 12--16, 2026, Karlsruhe, Germany}
\acmDOI{10.1145/3777912.3839787}
\acmISBN{979-8-4007-2327-8/2026/10}

\begin{document}

\title{Paid to Look Like Truth: The Prevalence and Dark Patterns of Advertorials in News Outlets}

\author{Emmanouil Papadogiannakis}
\affiliation{
	\institution{FORTH \& University of Crete}
    \city{Heraklion}
	\country{Greece}
}
\author{Panagiotis Papadopoulos}
\affiliation{
	\institution{FORTH}
    \city{Heraklion}
	\country{Greece}
}
\author{Nicolas Kourtellis}
\affiliation{
	\institution{Keysight AI Labs}
    \city{Barcelona}
	\country{Spain}
}
\author{Evangelos Markatos}
\affiliation{
	\institution{FORTH \& University of Crete}
    \city{Heraklion}
	\country{Greece}
}

\renewcommand{\shortauthors}{Emmanouil Papadogiannakis, Panagiotis Papadopoulos, Nicolas Kourtellis, Evangelos Markatos}

\ccsdesc[300]{Information systems~World Wide Web}
\ccsdesc[500]{Information systems~Online advertising}

\keywords{Advertorials; Online Advertising; Regulation; Dark Patterns; Web Monetization}

\input{sections/0.abstract}

\maketitle

\input{sections/1.introduction}
\input{sections/2.background}
\input{sections/3.methodology}
\input{sections/4.findings}
\input{sections/5.advertorials}
\input{sections/6.darkPatterns}
\input{sections/7.relatedWork}
\input{sections/8.conclusion}
\input{sections/9.acknowledgements}

\bibliographystyle{ACM-Reference-Format}
\balance
\bibliography{main}

\appendix
\input{sections/appendix}

\end{document}

%% file: preamble/preamble.tex
\usepackage{xspace}
\usepackage{booktabs}
\usepackage{graphicx}
\usepackage{multirow}
\usepackage{framed}
\usepackage{url}
\usepackage[T1]{fontenc}

%% file: preamble/macros.tex
\newcommand{\eg}{e.g.,\xspace}
\newcommand{\etc}{etc.\xspace}

\newcommand{\ie}{i.e.,\xspace}

\newcommand{\goodad}{benign ad\xspace}

\newcommand{\goodads}{benign ads\xspace}
\newcommand{\Goodads}{Benign ads\xspace}

\newcommand{\FirstEuLocation}{EU\#1\xspace}
\newcommand{\SecondEuLocation}{EU\#2\xspace}

\newcommand{\point}[1]{\par\noindent\textbf{#1}}

\newcounter{finding}
\renewcommand{\thefinding}{\arabic{finding}}

\newcommand{\finding}[1]{
  \refstepcounter{finding}
  \begin{leftbar}
  \noindent\textbf{Finding \thefinding:} #1  
  \end{leftbar}
}

%% file: sections/0.abstract.tex
\begin{abstract}
Advertorials represent a marketing strategy where advertisements are designed to resemble the style and tone of editorial content.
Despite their appearance, they are, in fact, paid content intended to promote a product, brand or service.
Studies indicate that advertorials are more effective (81\%) and less intrusive than traditional banner ads or pop-ups.
Despite regulatory efforts for clear disclosure of paid content, concerns persist about the deceptive nature of advertorials. 
Advertorials can mislead readers into believing that they are consuming unbiased editorial content.
In doing so, they gain undeserved legitimacy, by draping themselves in the credibility of the publication's design.

In this study, we conduct the first systematic large-scale study of advertorials. 
We propose a novel automated methodology for detecting problematic advertorials in the wild, and collect 186K ad URLs over a period of 5 months.
We investigate their prevalence and explore their structural and linguistic characteristics, finding that advertorials appear in 1 out of 3 news websites, including popular and credible outlets (\eg The Guardian, EuroNews, CNN).
Additionally, they often exhibit behavior associated with malicious practices and deliberately obscure or make legal disclaimers difficult to recognize, preventing users from identifying the promotional nature of the content.
\end{abstract}

%% file: sections/1.introduction.tex
\section{Introduction}
\label{sec:introduction}

In the digital age, advertising has become deeply intertwined with how users consume online content.
Articles, blogs, applications and media are provided free of charge precisely because the underlying content is funded by advertisements rendered by publishers on their platforms.
In recent years, native advertising has emerged as a response to the overly disruptive nature of traditional banner and display ads~\cite{mccoy2004study}, as well as the significant rise in Web ad-blocking, which is used by a staggering 42.7\% of internet users~\cite{adBlockingStats}.
Native advertisements are online ads purposefully designed to blend seamlessly with a website's content, frequently mimicking its structure and style.
Native advertising yields 53\% more views~\cite{sharethroughNative} and a $2-3\times$ higher Click-Through Rate (CTR) than traditional advertising~\cite{choozleCPM}. 
Furthermore, 85\% of internet users may find native ads more trustworthy than conventional formats~\cite{advertisingStatistics}. 

\begin{figure}[t]
    \centering
    \includegraphics[width=0.9\columnwidth]{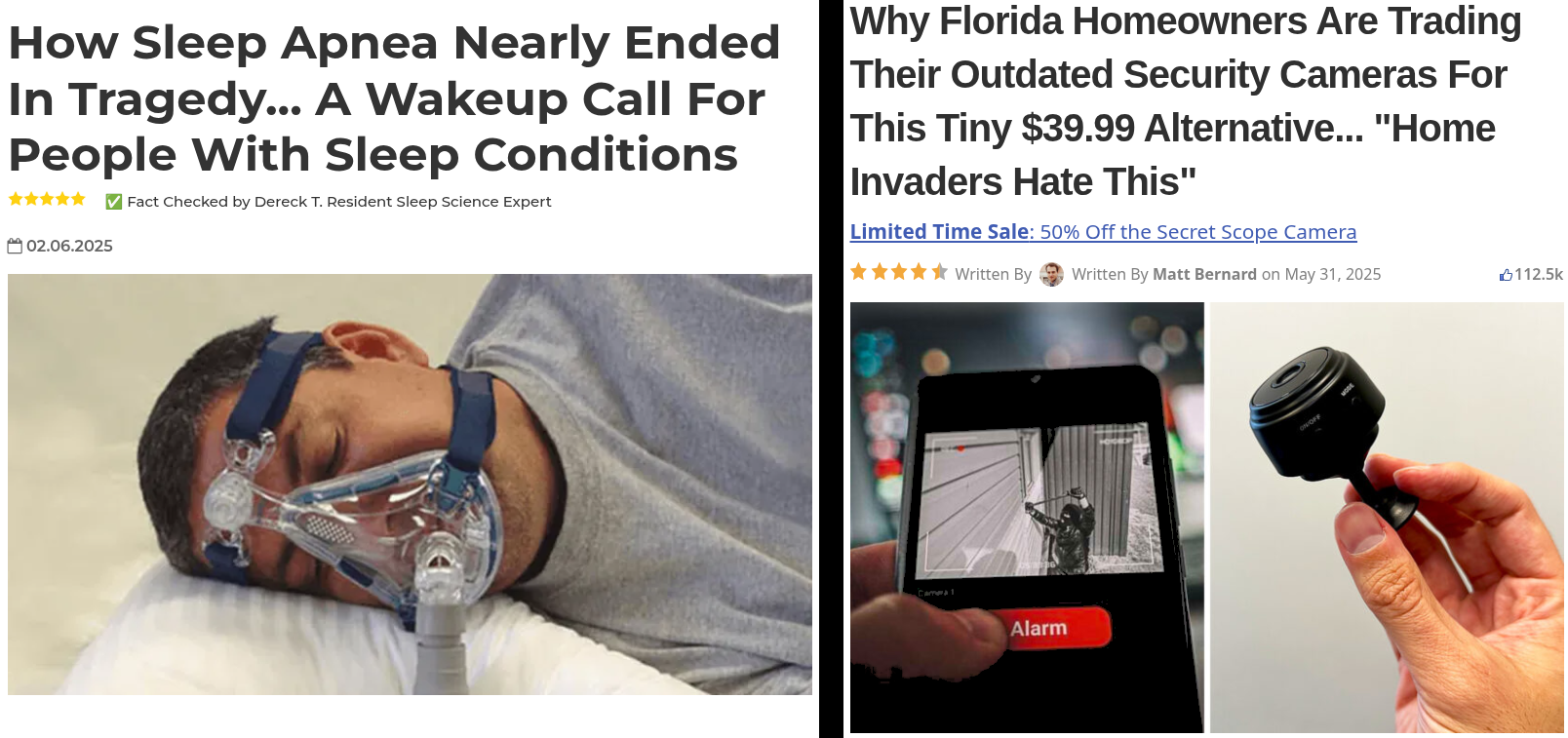}
    \caption{Example of advertorials; paid advertisements designed to look like an article, blending informative content with product promotion. Although they appear to present general information, they ultimately serve to promote a specific product (pillows on the left and cameras on the right).}
    \Description{An example of advertorials, which are paid advertisements designed to look like articles, blending informative content with product promotion.}
    \label{fig:advertorialExample}
\end{figure}

\emph{Advertorials} are a form of native advertising that blend editorial-style content with sponsored material and have been explicitly crafted to resemble the tone, format and language of genuine editorial articles.
Studies indicate that advertorials are 81\% more effective (and notably less intrusive) than traditional ads~\cite{effectiveAdvertorials}.
Consequently, Web users can sometimes unknowingly encounter content that appears to be a standard article but is, in fact, an advertisement promoting a specific brand, product or service.
Importantly, not all native ads lead to advertorials.
A native ad may instead direct users to a conventional product page, publisher content or another type of landing page.
In the case of advertorials, however, clicking a native advertisement on an ordinary website, may direct the user to an external landing page that presents promotional content in the style of editorial or journalistic content, as illustrated in Figure~\ref{fig:advertorialExample}.

The concept of blending advertising with editorial content is not new.
Early newspapers and magazines employed this format to bolster revenue without expanding traditional ad spaces.
However, it was the digital revolution that catalyzed the widespread growth of advertorials~\cite{advertorialDictionary}.
This proliferation is primarily driven by their higher engagement rates and lower intrusiveness.
In fact, studies confirm that 70\% of users prefer learning about a brand through blog posts and articles rather than traditional advertisements~\cite{articlesVsAds}.

Despite these benefits, the increasing difficulty in distinguishing between actual journalism and paid advertising makes advertorials a particularly problematic form, as they can mislead readers into unknowingly engaging with promotional content.
Users frequently fail to recognize promotional content embedded within editorial material, with only 9\% of people identifying such advertorials~\cite{amazeen2019reducing}. 
For instance, in 2013, The Atlantic published a story praising the Church of Scientology's leader, presented in the same format as its regular editorial content, which was later revealed to be a sponsored post~\cite{scientologyAdvertorial}.
This advertorial's focus on sensitive topics like religion highlighted the risks of poorly disclosed sponsored content.
More recently, various news outlets in Australia promoted gas energy without disclosing that these articles were sponsored by the fossil fuel industry~\cite{australiaAdvertorial}.
Compounding this issue, studies show advertorials can blend so seamlessly with website content, deceiving older adults~\cite{amazeen2020effects} or visually impaired people~\cite{kodandaram2023detecting}.
These issues raise concerns about how the inherent
lack of transparency in advertorials can influence public opinion and shape perceptions.

Various regulatory bodies, including the Federal Trade Commission (FTC), the European Commission, the Advertising Standards Authority (ASA) and similar organizations worldwide, have sought to curb the unrestrained use of advertorials and ensure that they are clearly distinguishable from editorial content~\cite{ucpd,ftc,asaCap}.
These efforts aim to protect users from sponsored content disguised as impartial journalism.
In the academic domain, while substantial research has focused on native advertising and problematic ads~\cite{zeng2020bad,ali2023problematic}, less attention has been paid to the deployment patterns of deceptive or misleading advertorials, as well as the extent to which they comply with existing regulations.

In this work, we conduct the first large-scale and systematic study of advertorials, investigating their prevalence, as well as their typical characteristics.
We set out to explore how frequently online users encounter \emph{problematic advertorials} that exhibit deceptive or potentially harmful characteristics and study the language, style and structure used by advertorial operators to mislead users.
We examine how advertorials employ common malicious tactics to defraud users, often mimicking financial scams, miracle cures and fraudulent campaigns, while also using dark patterns to bypass regulations.
Our study focuses on advertorial landing pages reached through advertisements displayed on news websites.
We focus on ads on news websites because advertorials are designed to look like news articles, which may make them more effective.
Through a systematic analysis of online ads, we (i) identify common structural and stylistic features of advertorials, and (ii) assess the frequency with which users are exposed to such content.
We then examine global regulations and identify systematic non-compliance by advertorial publishers.
Finally, we propose a novel methodology capable of detecting problematic advertorials at scale. 
To support further research, enhance transparency and help prevent user deception, we publicly release data used in this study~\cite{openSource}.

The contributions of this work include the following:
\begin{enumerate}
    \item {\bf Large-scale analysis}: We conduct the first large-scale and systematic study of deceptive advertorial landing pages across multiple countries, investigating their prevalence within the programmatic advertising ecosystem as well as their structural and linguistic features.
    
    \item {\bf Advertorial detection methodology}: We propose a novel automated approach for detecting problematic advertorials in the wild, leveraging both structural and semantic features. We apply this methodology to over 186K ad URLs collected over a five-month period.
    
    \item {\bf Global prevalence:} We discover over 10K instances of problematic advertorials served as ads in 193 news websites.
    We find that 1 in 3 news websites worldwide serve ads for advertorials, including some of the most popular and credible outlets, ranked among the top 2,000 websites globally.
    
    \item {\bf Geographical disparities:} Our analysis reveals clear regional differences in advertorial exposure, with U.S. users encountering $2.5\times$ more distinct advertorial domains than users in the EU. We find that European news websites are less likely to host advertorials, whereas those based in Asia or North America are 50\% more likely to feature such content.

    \item {\bf Content focus:} We show that advertorials contain more text than both \goodads and news articles, while having substantially smaller HTML footprints ($2.7\times$ smaller than \goodads) and interacting with fewer third parties. Among identified advertorials, 44\% address sensitive topics such as health-related products, services or supplements.
    
    \item {\bf Deceptive design patterns:} We identify dark patterns (\eg font size, color and text placement) that obscure legal disclosures, making them difficult to recognize. Disclaimers are often written in more sophisticated language than their surrounding content.
\end{enumerate}

%% file: sections/2.background.tex
\section{Background}
\label{sec:background}

\subsection{Advertorial Definition}
\label{sec:advertorial-definition}

An \textbf{Advertorial} is a form of advertising that has been purposefully designed to look and feel like editorial content.
That is, the tone, style and format of an advertorial resembles content written by journalists or writers to inform or educate their audience.
However, the true goal of an advertorial is to promote a specific product, brand or service.
Figure~\ref{fig:advertorialExample} presents representative examples of advertorial content, in which the headlines imply informative intent, such as offering solutions to specific problems, while the actual content is a paid promotion for a specific product.
The fundamental concept of an advertorial is to provide promotional content within an engaging narrative, a technique that dates back to the early 1900s~\cite{historyOfAdvertising,learningFromLegends}.
However, with the evolution of online advertising, advertorials have become extremely popular~\cite{advertorialPopularity}, as they are a perfect candidate for native advertisements because of their less intrusive nature. 

More formally, in~\cite{zeng2020bad}, product advertorials are defined as \emph{``Ads for consumer products written in the style of a blog post or news article that do not obviously disclose that they were written by the advertiser, other than in the fine print in the header or footer of the page''}.
A key characteristic is the journalistic tone, which leads readers to perceive it as a standard editorial piece, providing engaging content that retains the user's attention.
In addition, although various regulatory bodies require that advertisements contain a banner or disclaimer indicating that the content is sponsored, such disclosures are often in the footer fine print, making them easy to miss~\cite{zeng2020bad}.

In this work, we focus on {\bf problematic advertorials}: a subset of advertorials that exhibit characteristics commonly associated with deceptive or potentially malicious practices.
Such advertorials incorporate deceptive design practices (\ie dark patterns), misleading or fabricated content signals (\eg falsified testimonials or stolen images), indicators of security risks (\eg deceptive landing pages) or data exploitation mechanisms.
These characteristics are identified through observable features and security signals (Section~\ref{sec:methodology}) and align with known deceptive Web practices.

\subsection{Misleading Promotional Tactics}

In our setting, advertorials are reached through ads displayed on news websites.
More precisely, the user first encounters an ad (often in the form of native advertising) on a news website and, upon clicking it, is redirected to an advertorial landing page.
Rather than presenting a conventional product ad, this page is designed to resemble an editorial article while promoting a specific product or service.
We refer to this external landing page as an advertorial and consider it a distinct item from the ad.
The news website displays the ad that attracts the user's attention, whereas the advertorial is the promotional landing page reached after the user clicks on that ad.
Our analysis focuses on the \emph{advertorial domains} that host the promotional content rather than the ads displayed on the news websites themselves.
We also focus only on ads served on news websites, as advertorials are inherently designed to mimic the presentation and tone of news articles or other editorial content.

\begin{figure}
    \centering
    \includegraphics[width=0.85\columnwidth]{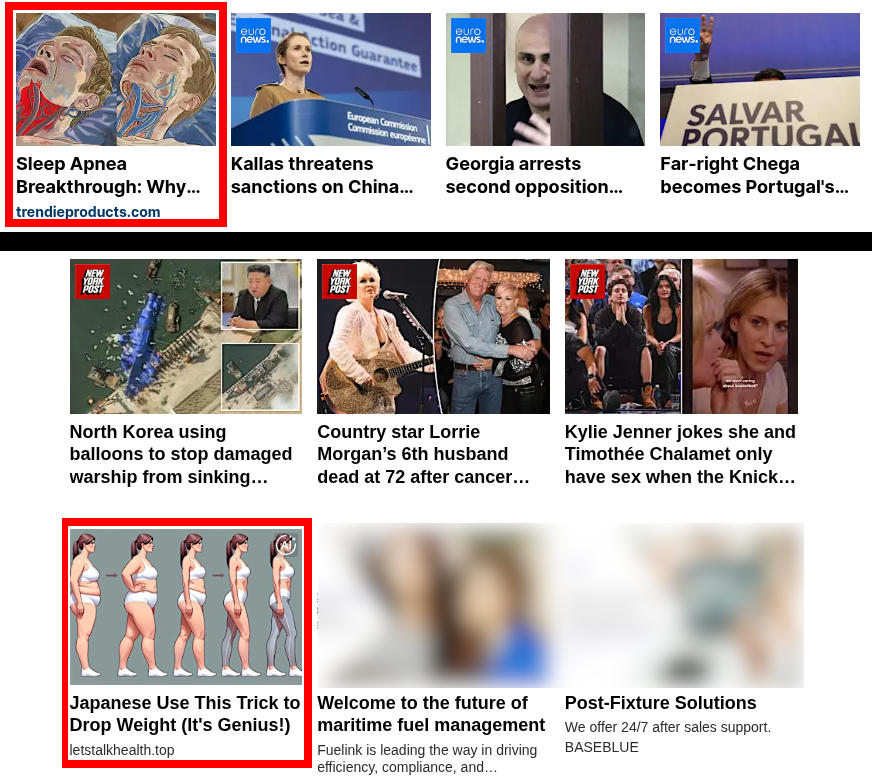}
    \caption{Advertorials (highlighted in red) being subtly promoted among articles of popular news websites. Ads are integrated into the news feed of legitimate news platforms, making them virtually indistinguishable from editorial content. By assimilating into the publication's design, they illegitimately acquire credibility.}
    \Description{Screenshot of a news website feed where sponsored advertorial articles are highlighted in red among regular editorial headlines. Ads are visually similar to legitimate articles. The visual design integration causes the advertorials to blend into the feed, giving them an appearance of editorial credibility.}
    \label{fig:nativeAdvertorials}
\end{figure}

Such ads are often served on popular and reputable news websites (\eg \emph{credible-news.com}).
As illustrated in Figure~\ref{fig:nativeAdvertorials}, the native ads are intentionally designed to blend seamlessly with the website's editorial content, making it difficult for users to distinguish them from actual journalism.
Although such content may be labeled (\eg as ``sponsored''), disclosures may be visually inconspicuous or cognitively overlooked (Section~\ref{sec:darkpatterns}), leading users to misinterpret ads as genuine editorial content.
When users click on the native ad, they are redirected to the advertorial hosted on \emph{health-product.com}.
There, the content typically presents misleading information in a format that appears factual and objective, often supported by fabricated reviews, testimonials or comments intended to convince consumers of the legitimacy and effectiveness of the promoted products.
A major concern is the authenticity of these testimonials and whether they originate from real consumers~\cite{grigsby2020fake,asquith2020critical}.
The above strategy constitutes an exploitative tactic that hijacks the trust people have in credible news websites.
Advertorials look like news articles, disrupting the user's advertising filters and skepticism, and people might assume that the ad is credible since it appears on a reputable website.
This tactic may also undermine the host website's credibility if users feel misled by the platform that hosted the native ad.
In this work we do not focus on paid editorial content published directly as part of a news outlet's own content.

\subsection{Regulation on Advertising}
\label{sec:regulation}

Various regulation frameworks regarding advertising aim to protect consumers from practices that may harm them or impair their ability to distinguish advertising from other content.
The European Union introduced the Unfair Commercial Practices Directive (UCPD) to protect consumers from deceptive marketing tactics~\cite{ucpd}.
The directive prohibits several misleading actions or omissions that could ``limit the consumer's ability to make an informed decision''.
Importantly, the directive explicitly classifies advertorials as misleading commercial practices, defining them as ``using editorial content in the media to promote a product where a trader has paid for the promotion without making that clear in the content or by images or sounds clearly identifiable by the consumer''.
Notably, ads must not deceive or mislead consumers, even if the information presented is factually correct.

\begin{figure*}[t]
    \centering
    \includegraphics[width=0.9\textwidth]{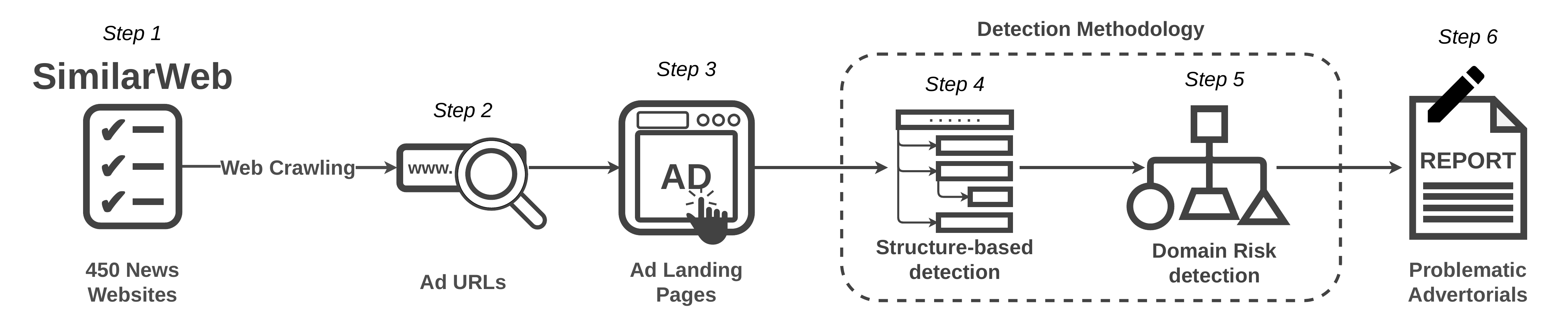}
    \caption{Overview of methodology for detecting problematic advertorials on news websites.}
    \Description{Diagram showing the methodology for detecting advertorials on news websites.}
    \label{fig:methodology}
\end{figure*}

Similar to the UCPD, the Federal Trade Commission (FTC) in the United States has issued guidelines that explicitly prohibit deceptive or unfair practices~\cite{ftc}.
They outline how to prevent deception and avoid misleading impressions by ensuring that disclosures are clear and prominently displayed.
In the United Kingdom, the Advertising Standards Authority (ASA) notes that advertorials are particularly difficult for users to recognize as advertisements due to their resemblance to editorial content, and therefore, must be explicitly disclosed as ads ``upfront, in a prominent and easily noticeable place''~\cite{asaCap}.
Likewise, the Australian Competition and Consumer Commission (ACCC) prohibits advertisements that mislead or deceive consumers, emphasizing that promotional content should include prominently displayed disclaimers~\cite{accc}.
The ACCC specifies that misleading conduct cannot be excused by disclaimers buried in fine print and explicitly prohibits the use of fake testimonials.

%% file: sections/3.methodology.tex
\section{Methodology \& Data Collection}
\label{sec:methodology}

In this work, we focus on problematic advertorials that use misleading or fabricated content, hidden disclaimers and manipulative visual design to obscure their sponsored nature. 
Such advertorials often use fake testimonials and images to give the appearance of legitimacy, further complicating detection.
Additionally, problematic advertorials often incorporate indicators of security risk, such as deceptive landing pages and bulk-registered domains.

To detect problematic advertorials and differentiate them from other forms of banner or native ads, we propose a novel data-driven automated methodology.
Our methodology relies on structural, linguistic and semantic features of advertorial text to discern it from legitimate editorial content or other forms of advertising (\ie \goodads).
We provide a high-level overview of the overall methodology in Figure~\ref{fig:methodology}.
After data collection and filtering is performed (Section~\ref{sec:data-collection}) we combine two complementary detection modules.
First, the landing page itself is analyzed to determine whether it exhibits structural characteristics associated with advertorials (Section~\ref{sec:structure-based-detection}).
This stage focuses on page properties and its presentation.
Second, we assess the suspiciousness of the associated domain using independent behavioral, reputational and technical signals obtained from external security and reputation services (Section~\ref{sec:domain-risk-detection}).
This stage operates at the domain level and assesses the potential risk associated with the hosting domain.
By combining the structural signal with the domain maliciousness signal we are able to identify problematic advertorials. 
An ad is considered a problematic advertorial only if both modules concur.
Finally, a manual evaluation (Section~\ref{sec:manual-evaluation}) assess the efficacy of our methodology.

\subsection{Data Collection}
\label{sec:data-collection}

To study advertorials, we first collect ads from websites in the wild.
Given that advertorials are intentionally designed to resemble journalistic content, we analyze advertorials reached through ads displayed on news websites.
News websites provide an important entry point for such ads because users may encounter native or recommendation ads while browsing news content and subsequently be directed to advertorial landing pages.
To that end, we compile a list of news websites from around the world, focusing on the ``News \& Media'' category provided by SimilarWeb~\cite{similarWeb} (Step 1 in Figure~\ref{fig:methodology}).
To capture geographic variations in the news advertising ecosystem, we prioritize a wider range across countries rather than sampling a larger number of websites from a smaller number of countries. 
In September 2024, we collect for each country the top five news websites, websites trending up in rank, websites trending down in rank, websites entering the top 100 and websites dropping out of the top 100 based on SimilarWeb data.
Our dataset covers the 59 countries for which SimilarWeb provides country ranking data.
This sampling strategy allows us to capture not only persistently popular news websites, but also news websites gaining or losing audience interest.
We note that the same news website can appear among the websites of multiple countries.
This collection process yields a list of over 1K news websites, out of which we retain only websites for which SimilarWeb also provides traffic and headquarters information, resulting in a final list of 450 news websites worldwide.
We make this list publicly available~\cite{openSource}.
We provide the distribution of their network traffic in terms of monthly visits in Figure~\ref{fig:trafficDistribution} ($x$-axis in log scale).
We observe that the list spans a wide range of audience sizes.
The estimated network traffic of the websites ranges from approximately 150K to 3.6B monthly visits (4 orders of magnitude difference), indicating that our list includes both small or local news websites and highly popular news outlets.
Thus, our dataset provides broad geographic coverage while capturing substantial variations in website popularity.

Using the open-source scraping tool of~\cite{papadogiannakis2022leveraging}, in Step 2 of Figure~\ref{fig:methodology} we visit the list of 450 news websites from three different user locations (\FirstEuLocation, \SecondEuLocation and U.S.) and collect all displayed ads. 
We use a VPN service to set user location (with two EU locations corresponding to different countries) and geolocation services to verify the crawler's location.
We visit each website with a clean browser state to keep its behavior and that of the advertising ecosystem independent of previously visited domains.
We also employ a browser extension~\cite{nouwens2022consent} to automatically interact with consent banners and accept all cookies by default.
As internal pages may differ from landing pages~\cite{10.1145/3419394.3423626}, we visit the landing page of each news website, as well as 5 internal subpages, selected via uniform random sampling.
Upon visiting a webpage, we wait for it to fully load, and then an additional 20 seconds to ensure complete ad transactions and rendering.
We then scroll down each page for 6 seconds (if possible) and then to the bottom to display all banner and native ads.
For each webpage, we collect its HTML content, the generated network traffic, cookie-jar (both first-party and third-party cookies) and detected ads.
To discover ads, we adopt the methodology of~\cite{papadogiannakis2023funds}, which extracts links from the rendered page and network traffic and evaluates them against the EasyList and uBlock Origin filter lists to determine whether they correspond to ads.
We extend this approach by additionally traversing all HTML elements in the rendered DOM and evaluating them against cosmetic filtering rules contained in the same filter lists.
For each element that matches a cosmetic filter, we examine whether the element (or an associated child element) contains a clickable URL.
If so, we classify the element as an ad and collect the respective URL.
We provide more information about the ad detection filter lists in Appendix~\ref{sec:ad-detection}.

We perform a repeated measures study and visit the entire list of news websites five times for each user location, totaling 15 distinct campaigns, from October 2024 to February 2025.
Altogether, we detect and process 186,124 distinct ad URLs, a considerable increase compared to previous work~\cite{zeng2020bad} ($\times$3 more ads collected during the crawling process and $\times$34 more ads successfully passed through the subsequent analysis pipeline).
Finally, we visit the collected ad URLs and analyze their landing pages (Step 3 in Figure~\ref{fig:methodology}).
We discuss ethical considerations of our work in Appendix~\ref{sec:ethics}.

\subsection{Structure-based Detection}
\label{sec:structure-based-detection}

Following the collection of 186K advertisement URLs, it is necessary to identify which ad landing pages correspond to advertorial content (Step 4 in Figure~\ref{fig:methodology}).
Advertorials have several characteristics that distinguish them from both traditional advertisements and editorial content, including legal disclaimers, user comments, testimonials and customer reviews.
We develop a structure-based method to detect such common advertorial characteristics.
Our methodology relies on both the website's structure (\ie HTML structure) and the legal text of disclaimers.

To build our system, we first assemble a formal collection of advertorial characteristics.
We manually visit and analyze the landing pages of 850 distinct ads found on news websites.
The ads are randomly selected from the entire set of all 186K distinct ads detected in Section~\ref{sec:data-collection}.
The goal of this manual inspection is to understand the structure of advertorials and identify HTML snippets that can be used for automated detection of advertorial landing pages.
The ads are not selected to cover specific news websites since we do not analyze how ads are placed on different news websites.
Instead, we focus on the landing pages that the ads lead to.

Next, on each ad landing page, we search for common patterns in advertorial landing pages across sampled ads and manually identify sections corresponding to testimonials, comments or reviews, and extract the HTML attributes that define these sections.
Our focus is on HTML class, ID and other HTML data attributes~\cite{mdnDataAttributes} (\eg \texttt{data-element-type}) that can serve as selectors to identify these specific sections on other websites.
For instance, we uncover that multiple advertorials have a specific user reviews section that can be singled out using specific HTML attribute. 
We also discover some fake comments on advertorial domains that explicitly resemble Facebook comments and can be identified using the attributes \texttt{class="comment-holder clearfix"} and \texttt{id="fbcomments"}.
We also consider the structural hierarchy of HTML elements, to identify precise instances of webpage sections and prevent misclassifications of unrelated pages that simply share similar attribute names.
These observations are incorporated into our detection pipeline (Step 4 in Figure~\ref{fig:methodology}) and each specific HTML attribute is tested against each ad landing page.

\begin{figure}
    \centering
    \setlength{\fboxrule}{1pt}
    \fbox{\includegraphics[width=0.9\columnwidth]{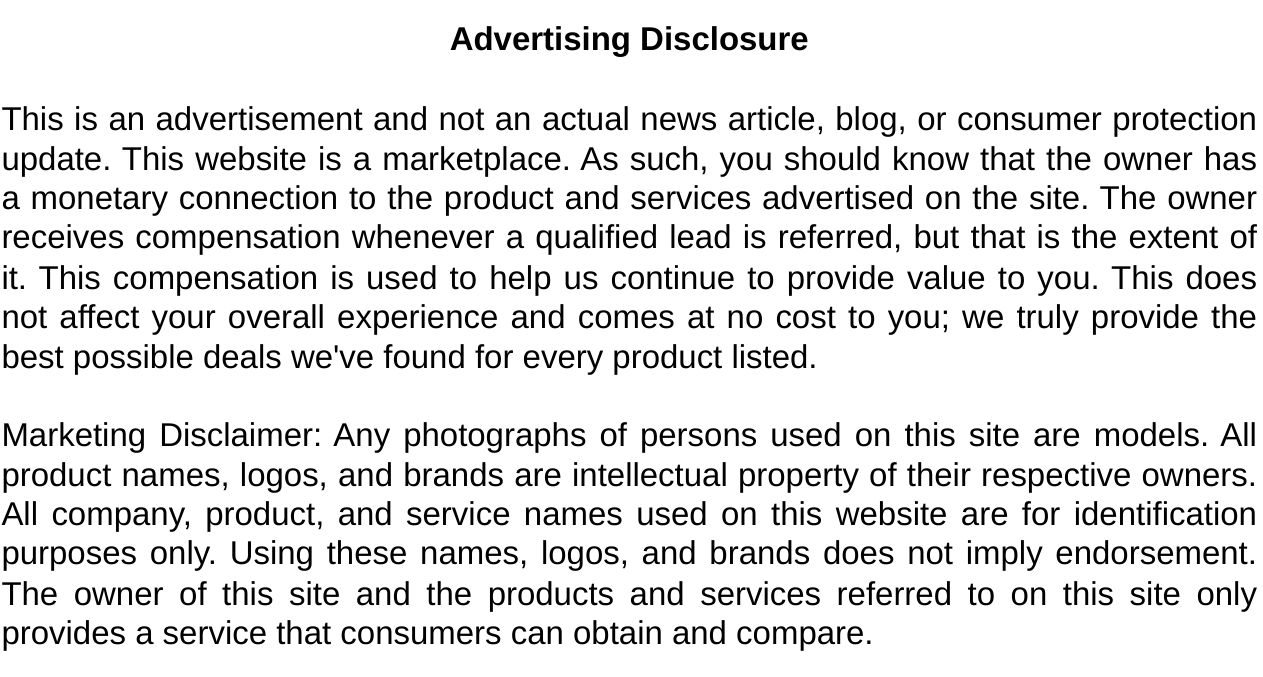}}
    \caption{Example of a legal disclaimer found on an advertorial website, indicating that the page is an advertisement.}
    \Description{Legal disclaimer example}
    \label{fig:disclaimerExample}
\end{figure}

For disclaimers that explicitly disclose an advertisement as an advertorial, we compile a list of legal vocabulary (\ie key phrases) that reveal the website's true nature.
An example of such a disclaimer is shown in Figure~\ref{fig:disclaimerExample}.
We further refine our system by collecting more detailed advertorial data. 
Specifically, we emulate different browser dimensions to detect responsive designs that present varied content across devices, and we use a VPN service to access websites from multiple countries, uncovering structural variations or disclaimers translated into different languages.
In total, we compile a list of 103 distinct legal disclaimer key phrases across seven languages (English, Dutch, German, Greek, Hebrew, Spanish and Portuguese) and make this list publicly available to support further research~\cite{openSource}.
We do not make use of machine translation as our detection methodology relies on exact matching of language-specific phrases and automatic translation can introduce inconsistencies or errors in the matching process.
We acknowledge that this list does not capture all possible disclaimers.
Our method provides a conservative estimate and may miss disclaimers that use alternative wording.

We develop a detection module that visits an ad landing page and extracts its text content and HTML structure.
This module tests a website against our dataset of disclaimer key-phrases and targeted HTML elements or page structure and classifies the URL based on whether its structure resembles the structure typically associated with advertorial content.
The module is then deployed on landing pages of the 186K ad URLs detected in Section~\ref{sec:data-collection} to assess whether they are candidate advertorials.

\subsection{Domain Risk Detection}
\label{sec:domain-risk-detection}

The structure-based detection methodology attempts to discover ads that resemble editorial content (\ie advertorials).
Next, to explicitly label an ad as a problematic advertorial, we evaluate whether it misleads, deceives or even defrauds users.
Relying on impartial Web engines, we build an automated classification system to assess whether the domains hosting these pages exhibit suspicious or malicious characteristics, based on behavioral and reputational indicators (Step 5 in Figure~\ref{fig:methodology}).
The advertorial landing pages in our dataset are hosted on domains different from the news websites on which the ads are displayed.
As a result, the domain risk analysis focuses on the advertorial providers rather than the news websites.

To build our automated advertorial filter, we first create a ground truth dataset containing 1,000 ads, randomly selected from the initial dataset of ads on news websites (Section~\ref{sec:data-collection}).
Then, we conduct a human annotation procedure to label instances of problematic advertorial content. 
This manual annotation procedure is independent of the one of Section~\ref{sec:structure-based-detection} and provides ground-truth labels for evaluating the resulting detection module.
Keeping these steps separate ensures that we do not design the detection rules using the same data that we later use to evaluate them.
The annotation phase involves 2 distinct reviewers that can label an ad as either an advertorial or a \goodad for legitimate display ads.
One of the reviewers is an author of this work, while the second is a knowledgeable (\ie computer science background) but independent (\ie not affiliated with this work) reviewer and is only informed about the advertorial definitions of Section~\ref{sec:background}.
Both reviewers are shown screenshots of ad landing pages and may browse them or use any translation tools they deem necessary.
Reviewers are not informed about where the ad was placed
and perform the annotation tasks independently, without interaction or shared decisions.

Following the manual annotation procedure, we analyze inter-rater agreement and find that reviewers classify 820 ads as benign and 157 as advertorials.
Reviewers disagree on only 23 ads, resulting in a 97.7\% agreement. 
The Cohen's kappa score is 0.92, indicating almost perfect agreement.
In the final dataset, an ad is labeled as advertorial only when both reviewers agree, thereby increasing confidence in the ground truth.
The resulting dataset is imbalanced, with more benign ads than problematic advertorials, reflecting real-world conditions in which benign advertisements outnumber advertorials~\cite{zeng2020bad}.
A brief follow-up interview conducted after the annotation task revealed that reviewers' decisions were primarily influenced by the presence of legal disclaimers and recurring structural patterns (\eg page layout, comment sections and customer review areas).
These features served as strong indicators for distinguishing advertorials.
Reviewers also noted that the editorial style of advertorials differs substantially from that of conventional ads. 
An illustrative example is provided in Appendix~\ref{sec:adsVsAdvertorials}.

\begin{table}[t]
\centering
\small
\begin{tabular}{lrrrrr}
\toprule
{\textbf{Target}} & {\textbf{Accuracy}} & {\textbf{Precision}} & {\textbf{Recall}} & {\textbf{F1}} \\
\midrule
Weighted Avg       & 0.928 & 0.929 & 0.928 & 0.923 \\ 
Suspicious Domains & 0.928 & 0.854 & 0.677 & 0.738 \\ 
Benign Domains     & 0.928 & 0.943 & 0.975 & 0.958 \\ 
\bottomrule
\end{tabular}
\caption{Domain risk classifier that achieves $\sim$93\% accuracy and F1 in detecting suspicious domains. 
The slightly lower performance on suspicious domains does not affect its role in the overall methodology.}
\label{tab:classifierPerformance}
\end{table}

The two detection modules of Figure~\ref{fig:methodology} address different properties.
The structure-based module identifies whether an ad leads to a page that exhibits the characteristics of an advertorial.
The second module operates at the domain level and estimates whether the domain hosting the landing page is suspicious or malicious based on independent behavioral, reputational and technical signals.
We use this domain signal as an additional indicator of problematic advertorials.
To that end, we build a classifier whose prediction target is domain maliciousness and not whether an ad is an advertorial.

To extract behavioral, reputational and technical data about domains, we build an oracle of 19 independent security, safety and analytics Web engines.
We select only reputable and trustworthy services, such as VirusTotal, Norton Safe Web and ScamAdviser.
We report the list of Web engines in Appendix~\ref{sec:engines-features} and note that the authors have no affiliation with these engines.
We also build a simple crawler that measures the size of the landing page in HTML format.
We make use of 20 signal sources (\ie engines) and extract 45 features per domain.
For training the domain classifier, the training instances are domains hosting the landing pages of ads that were manually labeled as either \goodads or advertorials.
To improve model efficiency and reduce overfitting, we apply feature selection using a random forest classifier.
Following common best practices, we encode categorical features as numerical values and impute missing data.
Feature importance is measured using Gini impurity and 24 features are retained, accounting for 96\% of total importance.
We discuss feature selection in Appendix~\ref{sec:engines-features}.

To build the detection module, we evaluate several classifiers (including K-Neighbors, Multilayer Perceptron, Gaussian Process) and find that a Random Forest Classifier performs best.
We split the labeled dataset of ad domains into training and test sets using an 80:20 random split performed at the domain level.
No constraint is imposed on the news websites on which the ads were served since the classification task focuses on the domains associated with the ads and not the source news websites.
We also perform hyperparameter tuning and cross validation with 20 iterations to accurately compute the model's true performance.
In Table~\ref{tab:classifierPerformance}, we demonstrate that the classifier achieves accuracy and F1 score of almost 93\%.
Although performance is slightly lower for suspicious domains, this is not concerning, as the classifier is only one component of the overall methodology.
We further discuss this in Section~\ref{sec:manual-evaluation}.
The proposed filter is applied sequentially after the structure-based detector (Section~\ref{sec:structure-based-detection}), providing increased certainty as to whether the landing page of an ad URL corresponds to a problematic advertorial.

\begin{table}[t]
\centering
\small
\begin{tabular}{lrrrrr}
\toprule
{\textbf{Target}} & {\textbf{Accuracy}} & {\textbf{Precision}} & {\textbf{Recall}} & {\textbf{F1}} \\
\midrule
Weighted Avg & 0.874 & 0.894 & 0.874 & 0.864 \\ 
Problematic Advertorials & 0.874 & 1.000 & 0.598 & 0.749 \\ 
\Goodads     & 0.874 & 0.845 & 1.000 & 0.916 \\ 
\bottomrule
\end{tabular}
\caption{Manual evaluation of the detection methodology on real and untested data. Detection accuracy is emphasized over dataset size to ensure reliability.}
\label{tab:methodologyManualEvaluation}
\end{table}

\subsection{Manual Evaluation}
\label{sec:manual-evaluation}

As illustrated in Figure~\ref{fig:methodology}, the proposed advertorial detection methodology consists of two sequential components: structure-based and domain risk detection.
We consider their combination essential, as the former identifies content with editorial structure, while the latter detects domains that may mislead or deceive users.
We group the two components into a filter that detects problematic advertorials (Step 6 in Figure~\ref{fig:methodology}) and evaluate its performance in the wild.
We run the entire pipeline on all 186K ad URLs and manually evaluate a subset of 1,000 URLs to measure its true performance.
These ads are randomly selected and have not been processed by the classifier at any stage (training or testing).
The classifier of Section~\ref{sec:domain-risk-detection} is trained and tested on a different set of manually labeled ads, allowing us to evaluate the detection pipeline on previously unseen ad landing pages.
The ad URLs correspond to 731 distinct ad landing pages, as the same ad may appear on different websites or visits.
We follow the same procedure as in Section~\ref{sec:domain-risk-detection} and manually annotate 502 ads as \goodads (68.67\%) and 229 as advertorials (31.33\%).

Then, we feed these 731 ads to our detection module and provide our findings in Table~\ref{tab:methodologyManualEvaluation}.
We demonstrate that our detection methodology achieves perfect precision in identifying problematic advertorials, albeit with lower recall.
This is a deliberate design choice.
A domain is classified as advertorial only when both the structure-based and domain risk detection modules concur.
Since this study focuses on domains serving advertorials, maintaining high confidence in detection accuracy is essential, even at the cost of a smaller dataset.
We argue that this conservative approach does not undermine the correctness of our findings, but rather it establishes a reliable lower bound on advertorial prevalence. 
This enables reliable conclusions about the nature and distribution of problematic advertorials, while avoiding incorrectly labeling legitimate editorial domains as advertorials.
We acknowledge that feature-based detection methods may be subject to evasion, and we therefore position our approach as a measurement tool for characterizing advertorials rather than a fully adversarially robust detection system.

%% file: sections/4.findings.tex
\section{Advertorial Prevalence \& Reach}
\label{sec:findings}

\begin{figure*}[t]
    \begin{minipage}[t]{0.32\textwidth}
        \centering
        \includegraphics[width=\columnwidth]{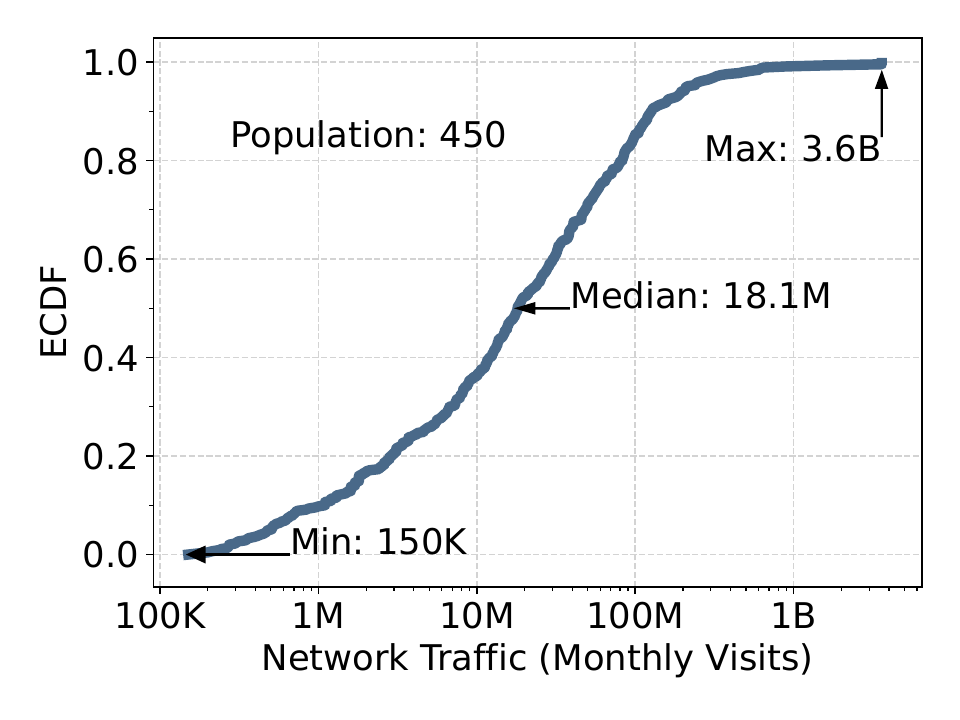}
        \caption{Network traffic distribution of news websites (monthly visits). The dataset contains both low-traffic (unpopular) and high-traffic (popular) sites, reflecting Web diversity.}
        \label{fig:trafficDistribution}
    \end{minipage}
    \hfill
    \begin{minipage}[t]{0.32\textwidth}
        \centering
        \includegraphics[width=\columnwidth]{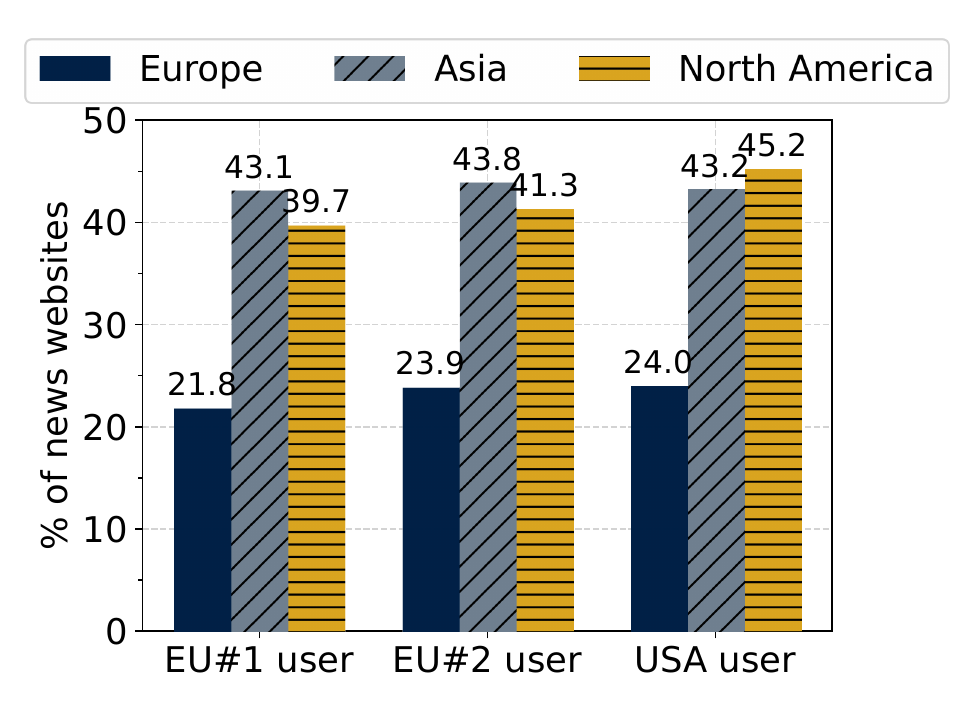}
        \caption{News websites containing advertorial ads, grouped by website location. News websites based in Asia or North America are $\sim$50\% more likely to contain advertorials.}
        \label{fig:websiteLocation}
    \end{minipage}
    \hfill
    \begin{minipage}[t]{0.32\textwidth}
        \centering
        \includegraphics[width=\columnwidth]{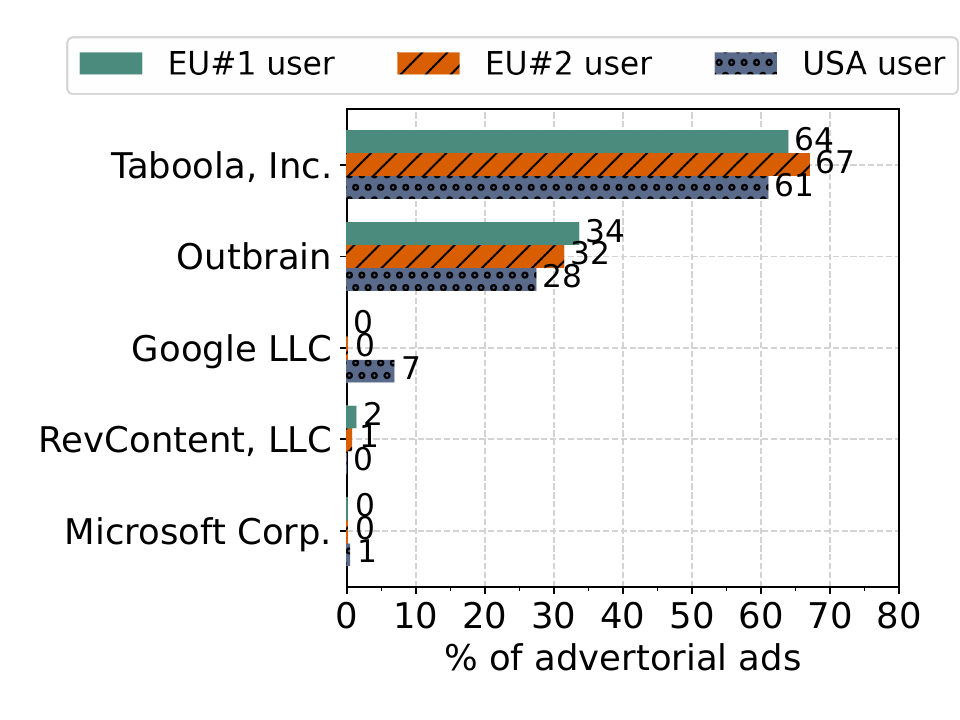}
        \caption{Ad Networks placing advertorial ads on news websites. Bars represent different user locations. Problematic advertorials are mainly promoted by Taboola and Outbrain.}
        \Description{Chart showing distribution of advertorial ads placed on news websites across different user locations. The chart highlights that problematic advertorial content is primarily promoted through the Taboola and Outbrain ad networks, which dominate the placement of such ads across multiple locations.}
        \label{fig:adNetworks}
    \end{minipage}
\end{figure*}

Using the data collection methodology described in Section~\ref{sec:data-collection}, we identify 186K ad URLs corresponding to more than 11K distinct ad landing pages, which represent the pages users reach after clicking on an advertisement.
We then input these domains into the detection methodology outlined in Figure~\ref{fig:methodology}, focusing exclusively on domains that trigger both the structure-based and domain risk filters.
This process yields over 10K ads found in 193 distinct news websites, directing users to 302 distinct domains that disseminate advertorial content.
We publicly release the list of advertorial domains~\cite{openSource}.

\point{User location:}
We examine the advertorials encountered by users located in different locations when visiting news websites.
A news website is considered to serve advertisements leading to problematic advertorials if our detection methodology identifies at least one such instance on either an internal page or the landing page during any visit.
Our findings, shown in Table~\ref{tab:userLocationResults}, reveal that the number of news websites containing advertorial ads is consistent across users from different locations, with only minor fluctuations.
In other words, there appears to be no significant difference in the news websites serving at least one advertorial to users within versus outside the EU.
Most importantly, we find that just over 1 in 3 news websites worldwide (between 34.7\% and 36.6\%) serve ads that may mislead or deceive their visitors (\ie problematic advertorials).

\begin{table}[t]
\footnotesize
\centering
\begin{tabular}{lrrrr}
\toprule
User & Crawled & \# news websites & \% news websites & \# advertorial \\
Location & Websites & with advertorials & with advertorials & domains \\
\midrule
\FirstEuLocation      & 435 & 151 & 34.71\% & 93  \\
\SecondEuLocation & 437 & 160 & 36.61\% & 88  \\
USA                  & 429 & 157 & 36.60\% & 223 \\
\bottomrule
\end{tabular}
\caption{Advertorials served across different user locations while visiting news websites. Websites containing advertorials are consistent across locations, but U.S. users encounter $2.5\times$ more advertorials than EU users.
}
\label{tab:userLocationResults}
\end{table}

Furthermore, we find that the number of domains serving advertorials is $2.5\times$ higher for users visiting from the United States than for users visiting from the European Union (last column of Table~\ref{tab:userLocationResults}), despite similar advertorial prevalence across news websites in the two regions.
This result highlights a geographical disparity in user exposure to advertorial content.
EU users tend to encounter the same advertorials across multiple websites, whereas U.S. users are exposed to a substantially larger and more diverse set of advertorials.
This increased exposure may hinder users' ability to distinguish between journalistic and sponsored content, potentially affecting unbiased decision-making.
We also examine the total number of ads served to EU and U.S. users and observe that the median number of ad URLs per news website is 36\% lower for U.S. users.
This finding indicates that the higher exposure to advertorials among U.S. users is not driven by an overall increase in advertising volume.

We acknowledge that the use of vantage points may affect ad delivery, with advertising platforms restricting ads from traffic associated with VPNs.
However, this behavior does not affect the correctness of our findings, but rather represents a conservative lower bound.
Finally, the observed differences in advertorial exposure between EU and U.S. users may be influenced by factors such as variations in regulatory pressure, enforcement and cultural differences in consumer protection.
Although our study does not directly examine these, future work could investigate regional differences in advertorial exposure.

\finding{
1 out of 3 of news organizations internationally serve ads for problematic advertorial content on their websites.
This is particularly evident for U.S. users that deal with $\times$2.5 more problematic advertorial domains compared to EU users.
}

\point{Location of operation:} 
Next, we investigate whether a website's location affects the presence of problematic advertorials on news websites.
While many news websites have a global presence and are visited by users from multiple countries, we focus on the location of each website's headquarters since editorial standards and business practices are often influenced by the hosting country.
Headquarters information is provided by SimilarWeb, as described in Section~\ref{sec:data-collection}. 
To better understand the effect of website location, we also group websites by continent using data from the United Nations Statistics Division~\cite{continentsDatabase}.

Our findings are illustrated in Figure~\ref{fig:websiteLocation}, focusing only on news websites from Europe, Asia and North America. 
Websites from other continents are excluded due to small sample sizes that do not ensure statistical representativeness (\eg only 11 websites from Oceania).
Every website is visited from each of the three vantage points (\ie \FirstEuLocation, \SecondEuLocation, U.S.).
We find that European websites contain the fewest advertorials, while websites hosted in Asia or North America are more prone to serving them. 
Approximately 23\% of European websites served an advertorial to their visitors, compared to roughly 43\% of Asian websites and 42\% of North American websites. 
Consequently, news websites hosted in Asia or North America are almost twice as likely to serve problematic advertorials to their users.
The consistency of these results across all user locations suggests that the findings are generalizable.
An examination of individual countries reveals that websites from India serve the most ads for advertorials, regardless of user location.

\point{Advertorial Prevalence in High-Traffic Sites:}
We further study the news websites that promoted at least one problematic advertorial to visitors in our data.
We find that ads for problematic advertorials appear even on some of the most popular and credible news websites worldwide, including The Guardian, EuroNews and CNN. 
Examples are provided in~\cite{openSource}.
Using the Tranco list~\cite{tranco}, which aggregates multiple sources to rank domains (ID: 998W2), we study the popularity of news websites promoting problematic advertorials.
We observe that 26.9\% (\ie more than 1 in 4) of news websites with advertorials are ranked in the top 2,000 worldwide.
Altogether, these websites receive an aggregated traffic of $\sim$11.8 billion monthly visits, increasing the risk of exposing vulnerable populations and impairing their ability to make informed decisions.
We also find no correlation between network traffic and advertorial presence (Spearman correlation = 0.1, $p$ = 0.006), suggesting advertorials appear on both popular and less popular news websites.

\finding{
Advertorials are found even on the most popular news websites around the world, with websites hosted in Asia and North America being more prone to serving advertorials.
In Europe, advertorials are more restricted.
}

\point{Advertorials Across Major Ad Networks:}
Given the complex nature of programmatic advertising, publishers often cannot fully control the type or quality of the ads displayed within their ad inventory.
As a next step, we examine the ad networks responsible for serving the ads that lead users to problematic advertorials.
For each ad leading to a problematic advertorial, we analyze the URL, extract its domain at the eTLD+1 level, and identify the operating ad network using the DuckDuckGo Tracker Radar dataset~\cite{tracker_radar}.
Domains without ownership information are excluded.
Figure~\ref{fig:adNetworks} shows that Taboola and Outbrain serve the majority of ads that lead to problematic advertorials ($\sim$64\% and $\sim$31\%, respectively) consistently across user locations, which is aligned with previous research~\cite{zeng2020bad}.
Notably, even major ad networks such as Google and Microsoft promote problematic advertorials. 
A manual analysis on a random sample of 100 ads from these networks confirmed that all were indeed problematic advertorials, showing that deceptive ads can bypass even the largest, most reputable ad networks.

\point{Repeated Exposure to Deceptive Advertorials:} 
Finally, we study the frequency with which certain advertorial domains appear as ads on news websites, identifying domains repeatedly displayed across dozens of sites. 
The domain \emph{ushoppyworld.com} appeared as an ad on 116 distinct news websites for the \FirstEuLocation user.
Similarly, \emph{web.expertmarket.com} appeared on 101 news websites for the \SecondEuLocation user.
We observe that users are repeatedly exposed to the same deceptive advertorials, often on credible, popular news websites, increasing the likelihood they will believe them (\ie manipulation through familiarity).
Additionally, we conduct a case-insensitive search for these advertorial domains in two of the authors' organization's unsolicited email inboxes and discover seven detected domains used the same advertorial narratives via spam emails.
This suggests that some operators attempt to reach users through multiple communication channels.
These channels are considered ``unwanted solicitations'' under UCPD and are explicitly prohibited.

\finding{
Ads for problematic advertorials are mainly served by Taboola and Outbrain while some advertorials are repeatedly displayed across numerous news websites and manage to bypass filters of even the largest ad networks.
}

%% file: sections/5.advertorials.tex
\begin{figure*}[t]
    \begin{minipage}[t]{0.32\textwidth}
        \centering
        \includegraphics[width=\columnwidth]{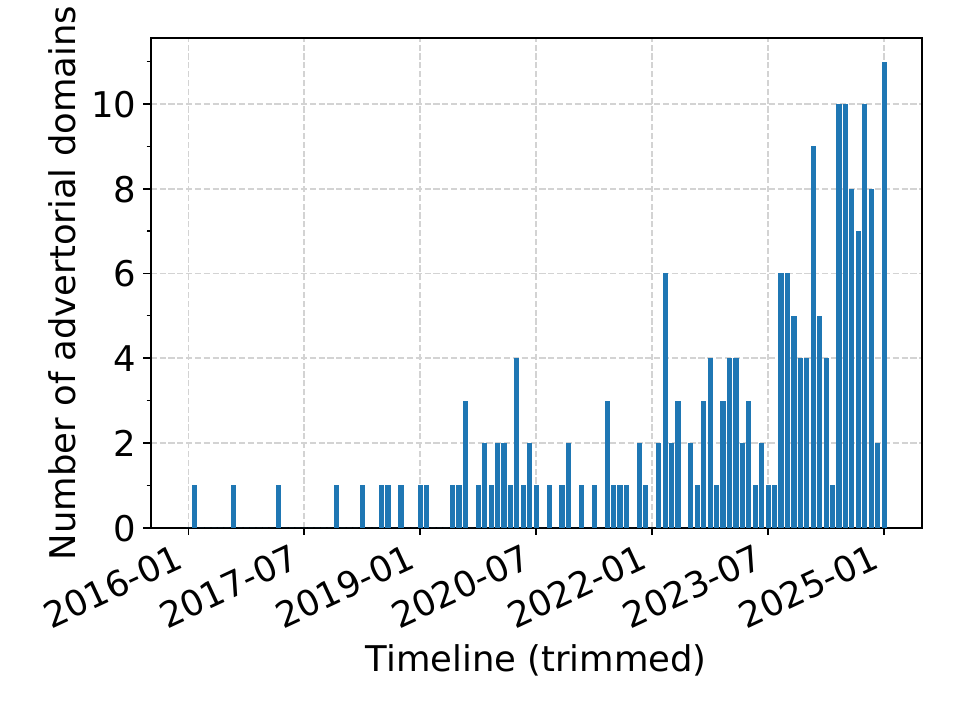}
        \caption{Registration date of advertorial domains. 56\% of these domains have been registered less than two years prior to data collection.}
        \label{fig:registrationDates}
    \end{minipage}
    \hfill
    \begin{minipage}[t]{0.32\textwidth}
        \centering
        \includegraphics[width=\columnwidth]{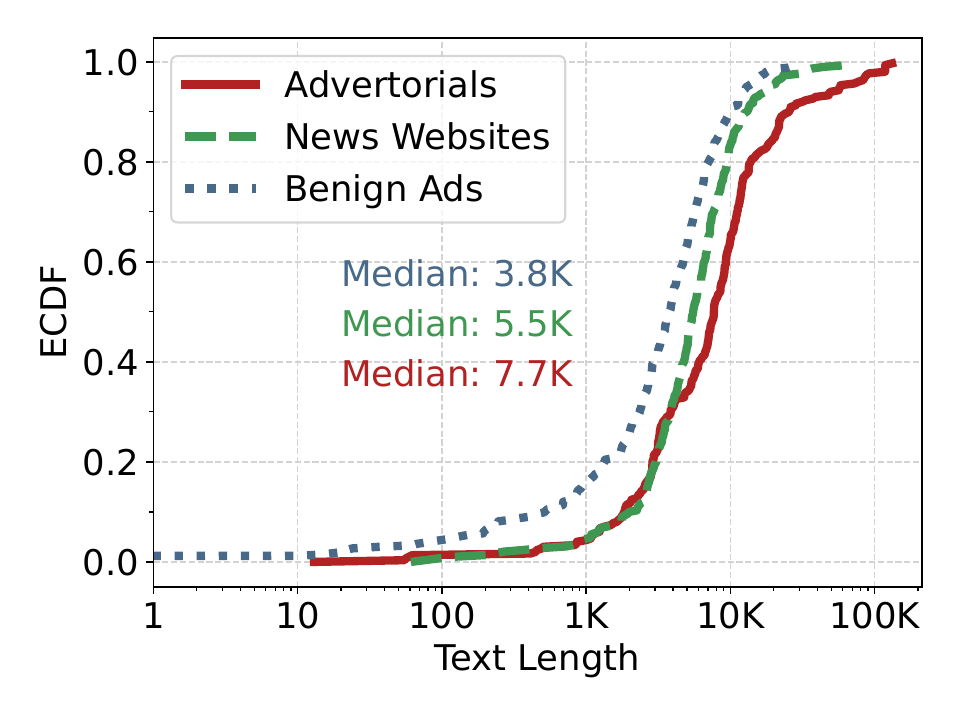}
        \caption{Distribution of text length for advertorials and other types of websites. Advertorials contain notably more text than \goodads and are closer in length to news articles.}
        \label{fig:lengthDistribution}
    \end{minipage}
    \hfill
    \begin{minipage}[t]{0.32\textwidth}
        \centering
        \includegraphics[width=\columnwidth]{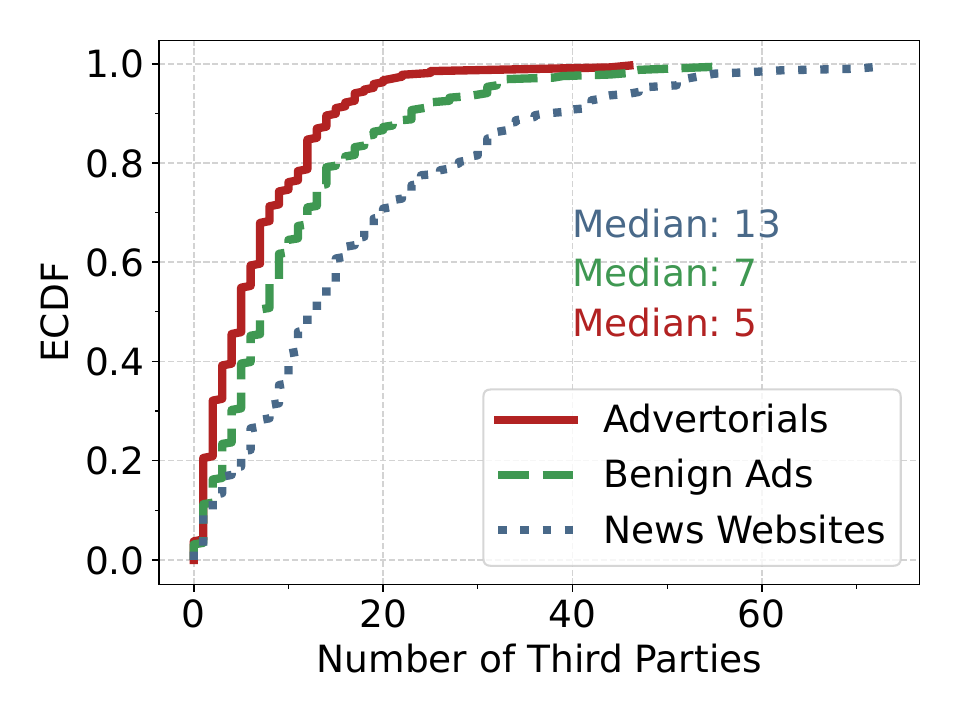}
        \caption{Distribution of third-party interactions across website types. Advertorial domains interact with fewer third-party trackers than \goodads and news websites.}
        \Description{Distribution chart (ECDF) comparing the distribution of third-party tracking interactions across different website types. Advertorial domains show fewer interactions with third-party trackers compared to legitimate \goodad websites and standard news websites, indicating a lower level of external tracking activity.}
        \label{fig:thirdPartyDistributions}
    \end{minipage}
\end{figure*}

\section{Characteristics and Content Analysis}
\label{sec:advertorial-characteristics}

\subsection{Domain Characteristics \& Lifespan}

We study the domains that host or distribute problematic advertorials to understand how such advertising operates at scale.
These domains may either be owned and operated by bad actors seeking to serve problematic advertorials, or be legitimate websites that have been compromised and unknowingly host problematic content, as observed in previous work~\cite{fittler2022prevalence}.
Given that both scenarios result in the distribution of similar advertorial content, we treat all domains equivalently. 
Using the Registration Data Access Protocol (RDAP)~\cite{rdap}, we successfully collect registration information for 213 problematic advertorial domains.
We observe that 81\% of them use the \emph{.com} TLD, rather than less familiar alternatives.
Following that, \emph{.org} and \emph{.shop} TLDs are less common, representing only 3\% and 2\% of advertorial domains, respectively.
Evidently, advertorial operators prefer the \emph{.com} TLD.
In contrast, fewer than 20 domains use less common TLDs such as \emph{.shop}, \emph{.online}, \emph{.club} or \emph{.biz}.

Next, we examine the registrars that registered the domain names.
We identify NameCheap as the most common registrar, with 59\% of problematic advertorial domains registered through it.
Notably, ScamAdviser reports that NameCheap has a high proportion of spammers and fraudulent sites, potentially due to a lenient registration process.
GoDaddy is the second most popular registrar, accounting for 21\% of domains. 
Together, these two registrars (responsible for 80\% of problematic advertorial domain registrations) appear to enforce advertising regulation policies ineffectively or offer only loose verification.
Both explicitly prohibit ``wrongful deception with the intent to gain a monetary benefit'' and ``misleading or deceptive content'' on registered domains.
Prior work has also shown that NameCheap and GoDaddy disproportionately register misinformation websites~\cite{han2022infrastructure}.

Finally, we examine the lifetime of advertorial domains using registration dates.
Results are shown in Figure~\ref{fig:registrationDates}, with a truncated $x$-axis highlighting recent years.
We find that most advertorial domains are relatively new.
More than half (56\%) are less than two years old at the time of writing (\ie registered after May 2023), and 88\% were registered after January 2020.
Newly registered domains can indicate low reputation or trustworthiness, or suggest malicious intent~\cite{akamaiNRDs}.
A closer inspection of domain names reveals evident patterns, such as \emph{wellnessguide103.com} and \emph{wellnessguide104.com}.
Bulk domain registrations over a short period may signal coordinated campaigns designed to evade detection or content-blocking systems.
Manual examination reveals cases in which different domains promote the same product, share identical structure and appearance, or even reuse the same user testimonials for different products.
We provide example screenshots in~\cite{openSource}.

\finding{
Advertorial operators mostly register recently created \emph{.com} domains, through registrars such as NameCheap and GoDaddy, with evidence of coordinated registrations.
}

\subsection{Advertorial Content \& Structure}
\label{sec:advertorial-content}

Next, we perform a deeper investigation into the content that problematic advertorial domains serve, that is, the text they use to persuade users to purchase a product or service.
A defining characteristic of advertorials is that they are written to resemble editorial or journalistic content.
Towards that, we compare their text length with that of other types of websites.

We construct a corpus of 300 problematic advertorials, 300 \goodads and 300 news articles based on data from the repeated-measures study (Section~\ref{sec:data-collection}).
\Goodads are randomly selected from the manual annotation process (Section~\ref{sec:domain-risk-detection}), while news articles are drawn from the pages visited by the crawler.
We randomly sample 300 instances from each category to obtain balanced and comparable sets, avoid differences in set size from influencing our analysis and maintain a manageable workload given our available computational resources.
We extract the full HTML and all visible plain-text content of each page and in Figure~\ref{fig:lengthDistribution} we plot the resulting text-length distributions.

We find that advertorials contain significantly more text than both \goodads and news articles.
Two-sample KS tests across all pairs of distributions show statistically significant differences, with $p$-values close to zero ($<1e^{-6}$).
Notably, advertorial text length is closer to news articles than to \goodads, with a KS statistic of 0.22 versus 0.37.
This supports the hypothesis that advertorials are deliberately crafted to mimic editorial content rather than traditional ads.
The logarithmic $x$-axis in Figure~\ref{fig:lengthDistribution} reveals substantial variation in advertorial text length.
That is, some advertorials are very brief, while others contain substantial content.
Further inspection shows that short-text advertorials primarily rely on embedded videos, whereas lengthy advertorials often cover multiple products.

Beyond the textual content presented to users, we also analyze the size of the full HTML pages.
Interestingly, although advertorials contain more on-page text than both \goodads and news articles, they exhibit the smallest HTML footprint.
The median advertorial page contains 106.3K characters: $2.7\times$ smaller than the median \goodad page (284.1K) and $4.6\times$ smaller than the median news article (484.8K).
This observation suggests that advertorial domains prioritize delivering user-facing text while largely avoiding complex page structures, SEO-oriented markup or integrations with advertising and tracking services. Such components typically inflate HTML size and introduce additional dependencies.

To assess this hypothesis, we quantify the number of third-party entities contacted during page loads.
We visit the corpus of 300 advertorials, 300 \goodads and 300 news articles, capturing all HTTP(S) traffic generated during the initial landing-page load.
We wait 10 seconds after page load to capture dynamic requests and automatically accept all consent banners to avoid blocking third parties.
Then, we process all requests and group their domains on the eTLD+1 level.
For example, requests towards \emph{region1.analytics.google.com} and \emph{accounts.google.com} are all grouped under the \emph{google.com} domain.
Figure~\ref{fig:thirdPartyDistributions} shows the number of third parties each website type interacts with.
We find that advertorial domains interact with fewer third parties than \goodads and news articles.
The median problematic advertorial website contacts just 8 distinct third parties, compared with 12 for \goodads and 19 for news articles.
When evaluating third-party domains against the Disconnect list of known trackers~\cite{disconnect}, we find that the median problematic advertorial domain interacts with 5 distinct third-party trackers.
This is still lower than the 7 trackers present on the median \goodad domain and the 13 trackers on the median news website.
Altogether, problematic advertorial domains focus on one-time visitors and avoid deploying third-party trackers that yield benefits for long-term or returning users.

\finding{
Operators of problematic advertorials prioritize the textual content of their webpages over complex designs or additional functionality.
They are often text-heavy and closely resemble news articles rather than \goodads.
}

\begin{figure*}
    \begin{minipage}[t]{0.32\textwidth}
        \centering
        \includegraphics[width=\columnwidth]{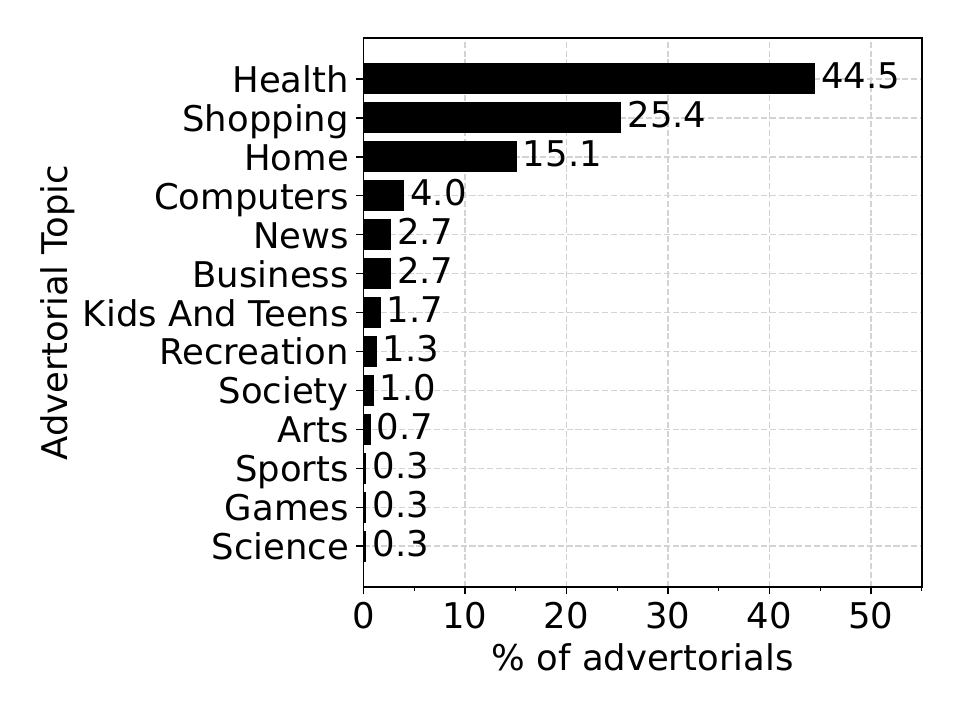}
        \caption{Distribution of advertorial topics in news websites. 
        Advertorials commonly promote health remedies and supplements, whereas this is comparatively rare in \goodads.}
        \label{fig:advertorialTopics}
    \end{minipage}
    \hfill
    \begin{minipage}[t]{0.32\textwidth}
        \centering
        \includegraphics[width=\columnwidth]{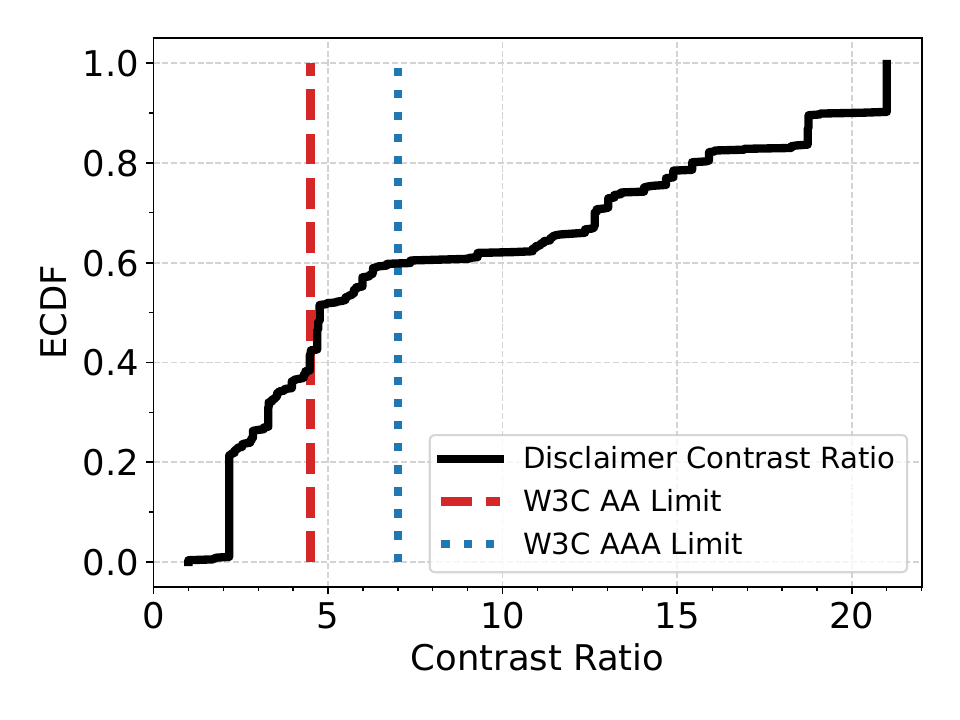}
        \caption{Advertorial disclaimers contrast ratio. The AA value is W3C's limit for accessible text. 42\% of disclaimers fall below this threshold, reducing visibility and readability.}
        \label{fig:contrastRatios}
    \end{minipage}
    \hfill
    \begin{minipage}[t]{0.32\textwidth}
        \centering
        \includegraphics[width=\columnwidth]{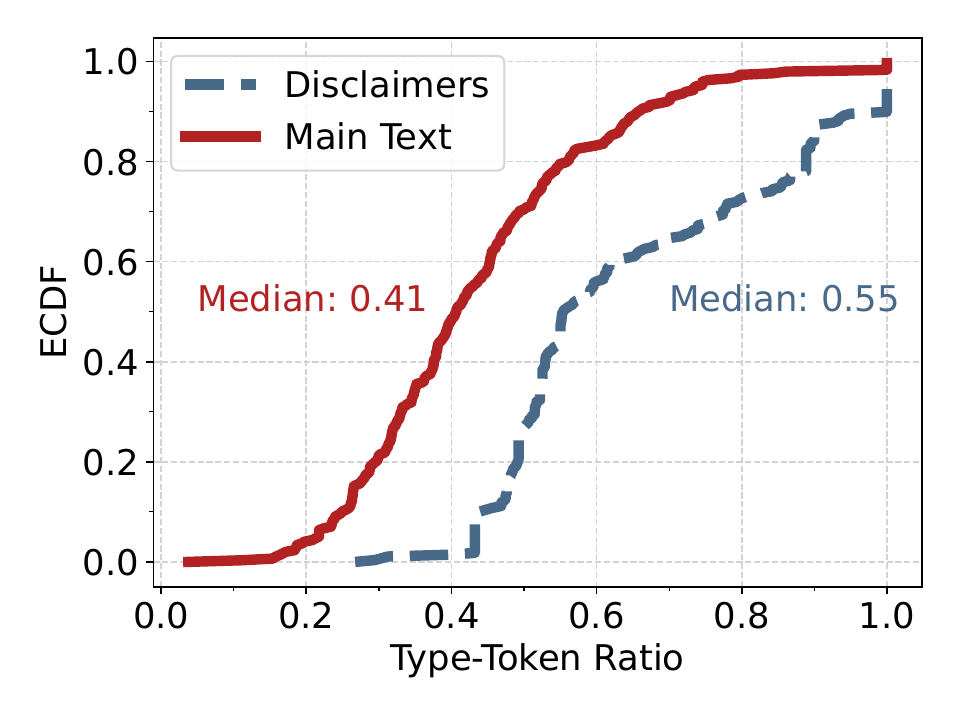}
        \caption{Type-Token Ratio scores measuring readability for advertorial text and disclaimers. Median TTR is 34\% higher in disclaimers, suggesting linguistic sophistication.}
        \Description{ECDF comparing Type-Token Ratio (TTR) scores between advertorial main text and associated disclaimers. The distribution shows that disclaimers have higher TTR values than main advertorial content, indicating disclaimers exhibit greater lexical diversity and linguistic complexity.}
        \label{fig:readabilityDistribution}
    \end{minipage}
\end{figure*}

Next, we analyze the thematic content of advertorials to identify the products or services being promoted.
We perform language-agnostic website classification using the pre-trained model of~\cite{lugeon2022language}.
We classify the landing page of each advertorial domain in its original language (\ie without translation) and incorporate visual features by providing a screenshot of the landing page to the classifier.
Each domain is assigned the category with the highest predicted probability, with Figure~\ref{fig:advertorialTopics} presenting the distribution of topics.

We find that the majority of advertorials (44.5\%) promote products or services related to Health, with Shopping and Home being the next most common categories.
Together, these three categories account for approximately 85\% of all advertorial topics.
Interestingly, \goodads rarely promote the Health category.
Instead, Business and Shopping dominate, each accounting for roughly 21\% of \goodads.
Health products rank only seventh, representing only 6\% of the \goodads analyzed.
Advertorials addressing sensitive topics such as health are particularly concerning, as they may pose systematic risks to user safety.
Indeed, the UCPD prohibits deceptive claims regarding the benefits or risks of products.

Next, we examine the languages in which advertorials are served.
We observe that advertorials are typically shown in English, followed by the user's native language.
Specifically, 52\% and 64\% of advertorials were in English for users in \FirstEuLocation and \SecondEuLocation, respectively, while 42\% and 33\% appeared in the primary language of the country from which the user endpoint accessed the website (\ie native language).
For users in the U.S., all advertorials were in English.
In a small number of cases ($<6\%$), the text appeared in a language associated with the location of the news website rather than the user.
Finally, some advertorial domains automatically translate their content based on the user's location.
We confirm this behavior using VPN connections from different geographic regions. 

\finding{
Advertorials most often refer to sensitive issues by promoting health products or services, while their text is usually in English.
}

%% file: sections/6.darkPatterns.tex
\section{Patterns of Malicious Behavior}
\label{sec:darkpatterns}

\subsection{Deceptive and Coordinated Activity}

Our analysis shows many advertorial domains display patterns associated with suspicious, malicious or illicit websites.

\point{Deceptive Landing Pages:} We observe that some advertorial domains have an empty or completely unrelated landing page.
When users directly type the advertorial domain name and navigate to it (rather than arriving from a third-party website), they encounter either a blank page or unrelated content.
We identify cases where the landing page corresponds to an architectural firm, a manufacturing company, a generic privacy policy, \etc, yet navigating to specific URLs within the same domain leads to problematic advertorials. 
In some instances, advertorial domains return a 404 error when visiting the landing page.
The domain \emph{ushoppyworld.com} displays a generic WordPress template as its landing page but serves multiple distinct advertorials on subpages. 
Altogether, we attribute such behaviors to attempts to conceal advertorial campaigns and evade detection, as documented by prior work (\eg~\cite{papadogiannakis2024ad,zhang2021crawlphish}).

\point{Content reuse:} 
Using perceptual hashing, we discover nearly identical images reused across different advertorials promoting the same product, including weight-loss results, product usage and customer profile photos.
We also identify ``authors'' publishing advertorials on multiple seemingly unrelated advertorial domains.
For example, one author (same name and profile picture) has published different problematic advertorials in at least 11 distinct domains. 
Another author has published an identical advertorial (title, content, images, structure) on the same day in 4 different domains, all of which respond with a $404$ error when users attempt to navigate to their landing page.
Manual verification shows that the listed authors are fabricated personas.
No verifiable social media or professional accounts could be linked to these ``authors'' and reverse image searches revealed that their photos appear on unrelated websites.
This behavior suggests that the domains may be controlled by the same entity or a coordinated network.
A similar pattern emerges in the ``user comments'' sections.
The same username appears across 7 problematic advertorial websites promoting weight-loss products.
These domains display identical comment threads with the same users in the same order and matching timestamps.
Even the user-submitted photos in ``testimonials'' are identical.
These patterns indicate fake accounts designed to mimic engagement or approval.

\point{Personal Data Harvesting:} We perform a manual analysis on a set of 10 problematic advertorial domains that receive the lowest scores on the security-engine features of our detection risk detection module (Section~\ref{sec:domain-risk-detection}).
Several of these domains are low quality and even flagged as scams or potential malware by security engines.
When users attempt to place an order for the promoted product, some domains collect sensitive personal information (name, address, email, phone number, \etc) under the pretense that ``a representative will soon get in touch.''
Notably, the domain \emph{ushoppyworld.com}, which exhibits multiple suspicious behaviors (\eg unrelated landing page, different advertorials per URL), also follows this pattern.
Several other domains present questionnaires that request personal details such as age, weight, height and address, supposedly to recommend the most suitable product.
However, regardless of the responses provided, the same product is promoted, revealing a deceptive and non-personalized marketing approach.

\finding{
Some advertorial domains exhibit a variety of suspicious patterns, including fake comments, fake authors, cloaked (hidden) pages, image re-purposing, data harvesting, \etc, that suggest malicious intent.
}

\subsection{Dark Patterns in Legal Disclaimers}

Advertorial operators may appear compliant by disclosing their intent, yet still violate regulations by mimicking news articles.
To assess this, we analyze disclaimer features in light of international regulations, with a particular focus on \emph{dark patterns}: design choices that subtly steer users toward actions that benefit the service provider~\cite{gray2018dark}.
We examine the legal wording that reveals the true purpose of the content (to promote a product or service) and differentiate between
(i) disclaimers, typically written in a formal style, explicitly describing the article's nature and the publisher's monetary gain, and
(ii) banners, usually a single word or short phrase, labeling an article as advertorial or sponsored content.
Figure~\ref{fig:disclaimerExample} shows an example disclaimer, while the most common banners are ``Advertorial'' or ``Sponsored Content''.
Our study focuses on the advertorial pages reached \emph{after} users click on ads displayed on news websites.
We do not evaluate how clearly these ads are labeled or distinguished from editorial content on the news websites themselves, as it is outside the scope of this work.
However, we acknowledge that since these ads are displayed through ad networks, they are typically labeled as ``Sponsored Content'' or ``Advertisement''.

\point{Disclaimer Content:} 
We examine the disclaimers published by problematic advertorial domains to disclose that the content is, in fact, an advertisement, even though it resembles editorial material.
These disclaimers are essential for advertorial operators, as they help protect their operations against legal claims.
We discover that the most common disclaimers are ``\textit{This website is an advertisement and not a news publication}'' and ``\textit{the owner has a monetary connection to the product and services advertised on the site}''.
This behavior is consistent for visitors located both within and outside the EU.
The disclaimers appear on 52\% and 45\% of advertorial domains, respectively, for \SecondEuLocation users and 43\% and 44\% for U.S. users.
Variations of these disclaimers are also common in legal disclaimers, for instance, the disclaimer ``\textit{This is an advertisement and not an actual news article}'' is found in 41\% of the advertorials the U.S. user visited.
Interestingly, 30\% and 25\% of the advertorials that the U.S. and \SecondEuLocation users came across, respectively, contain a very important disclaimer that ``\textit{This product is not intended to diagnose, treat, cure, or prevent any disease}''.

\point{Disclaimer Position:} 
We examine the locations of banners and disclaimers on advertorial domains.
We extract all text content from advertorial pages and compute
the relative position of each text element based on its coordinates in the rendered page layout.
We filter out text that is not visible, either because it is contained within hidden elements or because elements are placed outside the visible and scrollable portion of the page.
Our analysis shows that banners are typically placed at the top of the page, while disclaimers appear at the bottom.
The median banner is located at the top 0.06\% of the page (standard deviation 1.42), whereas the median disclaimer is at 96.12\% from the top (\ie at the bottom).
Advertorials do not prominently disclose their promotional nature, deceiving users and hindering their ability to recognize the content as advertising~\cite{amazeen2019reducing}.
Short top banners cannot convey all essential details and bottom-page disclaimers are often overlooked, leaving users unaware of critical information.
FTC guidelines require disclosures to be placed ``as close as possible to where users will look first''.
Similarly, the UCPD prohibits providing information in an ``untimely manner'' with users taking action before reading the disclaimer.
Advertorials fail these standards, placing disclaimers at the bottom of the page.

\point{Visual Presentation of Disclaimer:}
The FTC guidelines also state that ``disclosures should stand out, so consumers can easily read or hear them'' and that they should be ``in a font and color that is easy to read'' and ``in a shade that stands out against the background''.
We study whether these guidelines are properly enforced, focusing first on the visual presentation of text.
For disclaimers, we measure the Contrast Ratio, which quantifies the difference in luminance between text and background colors.
High contrast is necessary for readability~\cite{w3cContrastRatio}.
The contrast ratio ranges from 1:1, indicating no contrast (\eg black text on a black background), to 21:1, indicating maximum contrast (\eg white text on a black background).
The Web Content Accessibility Guidelines (WCAG) by W3C require a contrast ratio of at least 4.5:1 for normal text (AA) and 7:1 for accessibility to users with more severe visual impairments (AAA)~\cite{w3cContrastRatio}.
Figure~\ref{fig:contrastRatios} shows the distribution of contrast ratios for all detected disclaimers.
We find that the median disclaimer has a contrast ratio of only 4.76:1 and 41.92\% of disclaimers have a contrast ratio below the W3C minimum of 4.5.
This indicates that text and background colors are too similar, making the disclaimers difficult to notice and read.
The lowest contrast ratio we observe is gray text on a background of a slightly different shade of gray (Appendix~\ref{sec:dark-patterns-examples}).

\point{Disclaimer Text Prominence:}
Next, we evaluate whether disclaimers or banners are clearly identifiable by users and whether they stand out from other text on the page.
To this end, we extract all text elements from each advertorial page and compute their font sizes.
Analyzing the distribution of font sizes, we find that the majority of disclaimers and banners (69\%) have a font size smaller than the page's median.
As a result, disclaimers are deliberately smaller and harder to notice, effectively functioning as ``fine print''.
Surprisingly, only 5.5\% of disclaimers have a larger font size, indicating that only a small fraction stand out and can be easily identified.

\finding{
Even though advertorials contain legal disclaimers, they purposefully employ dark patterns to make these disclaimers hard to identify (\eg~small font size, faint font color, placed at the bottom of the screen).
}

Finally, we focus on the FTC guideline that disclaimers should be ``understandable'' and the UCPD requirement that information should not be presented in an ``unclear, unintelligible, or ambiguous'' manner.
We acknowledge that assessing text understandability is not trivial, as it depends on context and audience.
However, we employ two metrics to estimate how comprehensible and intelligible disclaimers are.
For each advertorial, we split its content into ``Main Text'', which contains the text promoting the product and written in journalistic form, and ``Disclaimer Text'', which contains the actual legal disclosure.
Text elements that do not constitute a legal disclosure are categorized as main text (\eg title, date, \etc).
For each type of text, we compute the Type-Token Ratio (TTR), which measures lexical diversity.
A high TTR indicates greater vocabulary diversity, often found in academic papers or literature, while a low TTR suggests repetitive vocabulary, 
typically found in informal speech or simple text.

Figure~\ref{fig:readabilityDistribution} shows the distribution of TTR values for disclaimers compared to main advertorial text.
We find that disclaimers have higher TTR values, with a median 34\% higher than that of the main text, suggesting that disclaimers are in a more sophisticated style than the rest of the article.
To verify this, we also compute the Linsear-Write metric, which measures readability of technical texts.
This metric helps assess how accessible advertorial content is to general audience.
We find that the median disclaimer text has a score of 8.57, while the median main text scores 5.6.
These correspond to an 8\textsuperscript{th} or 9\textsuperscript{th} grade reading level for disclaimers, compared to a 5\textsuperscript{th} or 6\textsuperscript{th} grade for main content.
Consequently, the main advertorial text is easier to read and more accessible to a general audience than the more complex disclaimer text.

\finding{
The legal disclaimer text in problematic advertorials is more sophisticated than the rest of the website text and may not be understandable by a general audience.
}

%% file: sections/7.relatedWork.tex
\section{Related Work}
\label{sec:related}

Prior studies have examined online advertising ecosystems, including ad targeting, bidding strategies and delivery optimization, highlighting how mechanisms influence what ads users are exposed to (\eg~\cite{10.1145/3517745.3561450,10.1145/3517745.3561460}).
Blending advertising with content is a long-standing practice predating advertorials.
Research shows users often struggle to distinguish Google Search ads from organic results due to unclear labeling~\cite{lewandowski2018empirical}, leading to unintentional clicks.
Similarly, native ads in large language models (LLMs) are highly effective and well-perceived when contextually relevant~\cite{zelch2023commercialized}. 
A study of 800 adults~\cite{amazeen2020effects} found only 9\% recognized native ads, with recognition leading to decreased trust in publishers. 
While health-focused advertorials are well-documented~\cite{kim2017advertorials}, they are generally more effective than traditional ads at manipulating purchase intent while remaining unobtrusive.
Content Recommendation Networks that distribute sponsored links through widgets such as ``Recommended For You'' have also been studied, showing that they often fail to clearly disclose the paid nature of content and highlighting the distribution of ad content that can resemble editorial recommendations~\cite{bashir2016recommended}.

In~\cite{10349218}, the authors discussed the prevalence of problematic ads within Virtual Reality (VR) applications, which may require users to perform controversial activities (\eg simulating drug use) or subject them to distressing experiences (\eg depiction of death or torture).
They emphasized the necessity of explicitly labeling VR advertisements to prevent deception and facilitate informed user decisions.
In~\cite{huang2017detection}, the authors developed a deep learning system to detect misleading statements in advertisements for illegal cosmetics, food and drugs.
Similarly, a neural network that detects deceptive ads by analyzing text, figures and HTML structures we developed in~\cite{hsien2017data}.
In~\cite{zeng2020bad}, the authors found that most problematic ads on news and misinformation websites originate from native advertising.
Unlike our large-scale systematic analysis, their manual study used a significantly smaller dataset (34$\times$ fewer ads) to compare display and native formats.
Both studies identified Taboola as the primary source of problematic content, with Google contributing minimally.
A study of children's websites~\cite{10646733} identified over 1K improper ads (\eg weight-loss and sexually suggestive ads).

Research also indicates that Google ads on Android can expose users to malware~\cite{ma2024careful} while fraudulent websites use Google and Facebook to target their victims~\cite{bitaab2025scammagnifier}.
Studies show general user hostility toward intrusive ads, particularly those that ``blend in'' or promote ``scammy'' health supplements~\cite{zeng2021makes}.
``Creepy'' targeted ads further drive demands for data transparency~\cite{10179452} and a large-scale Facebook study confirms widespread user dislike for deceptive or sensitive content~\cite{ali2023problematic}.
Moreover, an analysis of 9K UK Advertising Standards Authority complaints revealed that misleading ads dominate the commercial sector~\cite{auxtova2025offensive}.
Similarly, a study of 10K websites found 10\% non-compliant with Acceptable Ads Standards due to intrusive formats like pop-ups and autoplay videos~\cite{zafar2025assessing}.
Furthermore, ad disclosures are frequently non-compliant on Facebook~\cite{9152626} or absent on YouTube and Pinterest~\cite{mathur2018endorsements}.

Similar to our approach, the authors in~\cite{8418597} developed an automated detection system for online survey scams by analyzing website structure, content and word sequences.
They found that deceptive ads can lure users toward malicious survey gateways and then direct them to scam pages. 
Using a comparable methodology,~\cite{kotzias2023scamdog} investigated e-commerce scams.
Much like our work, they trained a Random Forest classifier using HTML features, though they also incorporated robots.txt, network, HTTP and domain-level features.
Furthermore, utilizing telemetry data from millions of devices, a study of 607,000 scam domains revealed that ``Shopping'' is the most prevalent scam category~\cite{kotzias2025CtrlAltDeceive}.
Notably, 13.3\% of users reached these domains by following an advertisement, a behavior similar to that observed with problematic advertorials. 
Scam campaigns enticing users with gifts or investment opportunities are not limited to advertorials, as they also appear within the comments sections of various online platforms~\cite{li2024like}.

%% file: sections/8.conclusion.tex
\section{Discussion \& Conclusion}

\point{Discussion:}
Problematic advertorials blur the line between journalism and advertising, intentionally deceiving users.
Concerns exist regarding whether the average consumer (or demographic groups) can recognize them as paid promotions rather than impartial articles.
This misunderstanding can shape consumer opinions, reducing their ability to make informed decisions. 
Furthermore, advertorials are often hosted within coordinated networks to evade detection, while showing suspicious, potentially malicious behavior.
Our findings highlight that there are significant weaknesses in the practical protection of users by existing disclosure regulations.
We note that in this work we do not distinguish between 
limitations in the regulations themselves and insufficient enforcement of those regulations.
Geographical differences in exposure suggest that regulatory frameworks and cultural factors shape user experiences, underscoring the need for region‑specific interventions. 

\point{Recommendations:}
To improve the detection and mitigation of deceptive advertorials, monitoring should extend beyond low-traffic or suspicious websites.
Advertising platforms, publishers and measurement systems should treat advertorial detection as a general problem rather than an issue limited to obscure publishers.
We believe the methodology presented in this work can help towards this direction.
In practice, monitoring systems can periodically crawl ad landing pages and flag cases in which the destination resembles editorial content but contains promotional material.

Moreover, detection systems should combine multiple signals instead of attempting to identify problematic advertorials from visual appearance or legal disclosures only.
We demonstrate that structural features and the presence of specific components like user comments can be used to distinguish ordinary advertising landing pages from pages designed to imitate editorial content.
Additionally, the use of recently registered domains along with the relatively small HTML footprint and limited third-party functionality of many landing pages can contribute to a risk score that determines whether a landing page receives additional analysis.
Advertorials related to health and disclosures that are difficult to understand should be considered high risk cases.
Since health products are often promoted through advertorials (Section~\ref{sec:advertorial-content}), advertising and publisher review processes could apply stricter scrutiny to campaigns in this category, particularly when claims are presented in an article format.
Disclosure checks should also evaluate not only whether a disclaimer is present, but whether it is sufficiently prominent and understandable relative to the surrounding content (Section~\ref{sec:darkpatterns}).

To address the challenges posed by advertorials, a combination of policy, technological and community-driven measures is required.
Clearly defined and better-enforced regulations can have a significant impact, while increased commitment from the advertising ecosystem is needed to prevent ads from appearing on legitimate websites.
Search engines, browsers and social media platforms can contribute by improving filtering, annotation and warning systems to reduce user exposure.
Platforms like VirusTotal (see Appendix~\ref{sec:engines-features}) can also support automated flagging and mitigation strategies.
Due to the prevalence of online advertising and the tools that filter it~\cite{10.1145/2815675.2815705}, enhanced ad blockers or ML-based browser plugins that automatically detect problematic advertorials can further increase transparency and reduce exposure to hidden advertising. 

\point{Future Work:}
This study highlights several promising directions for future work.
First, a longitudinal analysis could examine how advertorials evolve over time, including their lifetime, disappearance and reappearance and potential migration to new domains.
Such an analysis could also investigate how advertorial practices change in response to regulation or novel detection mechanisms.
Second, future work could examine advertorials and paid editorial content hosted directly on news websites.
While ad-blocking tools can reduce exposure to many forms of online advertising, advertorials are often indistinguishable from regular articles at the structural level, while publishers often deploy anti-adblocking mechanisms~\cite{10.1145/3131365.3131387}.
As a result, they might not be consistently addressed by traditional ad-blocking mechanisms.
Evaluating the extent to which existing ad-blocking mechanisms mitigate problematic advertorials would provide significant insight.

\point{Conclusion:}
Advertorials are ads intentionally designed to resemble editorial articles, subtly promoting products without the user's awareness.
In this work, we perform a large-scale, systematic study of advertorials that draw user traffic from ads served on news websites.
We propose a novel detection methodology leveraging structural, semantic and linguistic features, which accurately detects problematic advertorials. 
We find that advertorials are both widespread and frequently published, with $\sim$35\% of news websites worldwide serving at least one ad for a problematic advertorial domain. 
We also uncover that advertorial domains often exhibit multiple suspicious behaviors, such as fake comments, fabricated authors, hidden pages, image reuse and data harvesting, collectively suggesting malicious intent.
Finally, our findings highlight inconsistent adherence to international regulations, with legal disclaimers being purposefully difficult to recognize due to their size and placement, while also containing sophisticated text.
Overall, our work emphasizes concerns regarding the effectiveness of regulations and underscores the need for stronger consumer protection.

%% file: sections/9.acknowledgements.tex
\begin{acks}
Funded by the European Union (ATHENA grant ID 101132686 and C-SOC grant ID 8515915). Views and opinions expressed are however those of the author(s) only and do not necessarily reflect those of the European Union, European Research Executive Agency (REA), or the European Cybersecurity Competence Centre. Neither the European Union, the granting authority, nor the European Cybersecurity Competence Centre can be held responsible for them.
\end{acks}

%% file: sections/appendix.tex
\section{Ad Detection Filter Lists}
\label{sec:ad-detection}

We utilize the filter lists from EasyList and uBlock Origin for ad detection. 
Following the methodology of~\cite{papadogiannakis2023funds}, we use these lists to identify ads in URLs extracted from rendered pages and network traffic.
We additionally extend this approach to identify ads based on their representation in the rendered DOM.
We download both lists on October 7, 2024 (EasyList commit \texttt{3835a34}, uBlock Origin release 1.60.0).
We use filter list versions that differ from those employed in~\cite{papadogiannakis2023funds} in order to follow the latest ad rules. 
The filter lists contain both network filtering rules and cosmetic rules (also referred to as content rules).
Cosmetic rules identify page elements that should be hidden or modified based on properties of the HTML element, such as its ID or class, as well as CSS selectors.
As in~\cite{papadogiannakis2023funds}, we make use of the Brave ad-blocking engine~\cite{braveAdblock} to enforce these rules.
Although EasyList and uBlock Origin use slightly different syntactic conventions, their filtering rules follow largely similar syntax and are parsed and applied by the ad-blocking engine.
Finally, we respect and apply exceptions for specific domains defined by the filter lists, ensuring that rules are not applied to unrelated domains.

For cosmetic filtering, we traverse the entire DOM tree of each rendered page and test each element against the cosmetic rules from the EasyList and uBlock Origin filter lists.
For each element that matches a cosmetic rule, we examine whether the element or an associated child element contains a clickable URL.
If so, we classify the element as an ad and collect the corresponding URL.
This complements the network and URL-based detection employed in~\cite{papadogiannakis2023funds} by allowing us to detect ads whose presence is identifiable from the rendered page structure.

\section{Ethics}
\label{sec:ethics}

This research was conducted in accordance with recognized ethical standards and protocols for information studies~\cite{kenneally2012menlo,rivers2014ethical}.
In compliance with the GDPR and ePrivacy regulations, we do not collect or process personal information of real users.
We purposefully configure our crawling systems to emulate the behavior of real users, ensuring that all automated actions are rate-limited and unobtrusive.
Our system waits until the page fully loads, scrolls and navigates through the website at a natural pace, and waits for up to one minute before proceeding to the next webpage.
Finally, we make a conscious effort to avoid impacting the advertising ecosystem.
We only click on each ad URL only once, and considering the thousands of distinct advertisers we identify (Section~\ref{sec:findings}), the 5-month duration of our analysis (Section~\ref{sec:data-collection}), and the low cost per thousand ad impressions~\cite{pachilakis2021youradvalue}, we believe our activity did not meaningfully deplete any advertiser budgets.
This approach is consistent with prior work that has employed similar ad-clicking methodologies to study advertising ecosystems.
\cite{zeng2021polls} acknowledges the potential for minor financial impact due to impression- or click-based pricing models, but empirically estimates the resulting costs to be negligible at scale, particularly when distributed across a large number of advertisers.

\section{Advertorial Linguistic Tone}
\label{sec:adsVsAdvertorials}

\begin{figure}[t]
    \centering
    \setlength{\fboxrule}{1pt}
    \fbox{\includegraphics[width=0.8\columnwidth]{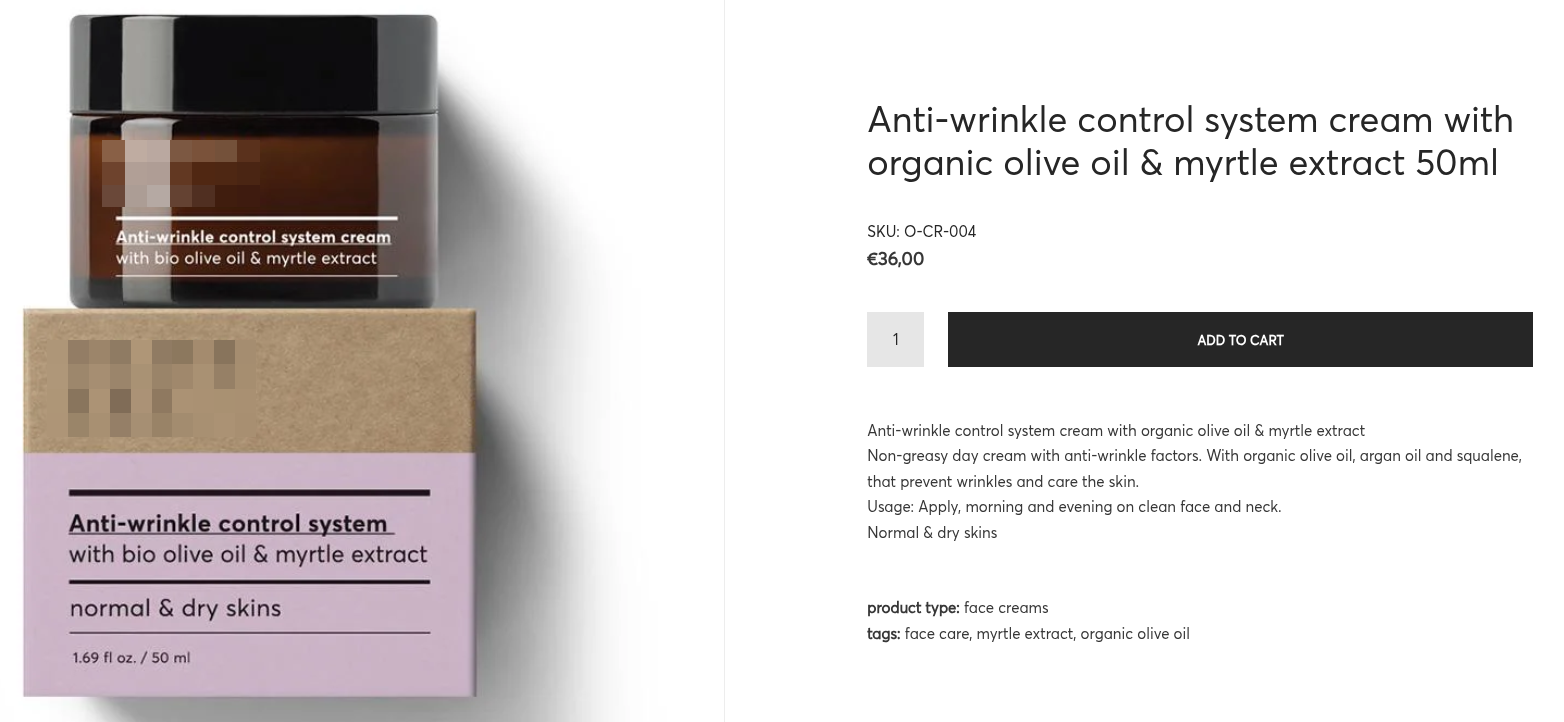}}
    \caption{Example of a traditional anti-wrinkle cream ad using a direct persuasive approach, highlighting benefits and encouraging purchase with concise text.}
    \Description{Example advertisement for a traditional anti-wrinkle cream presented in a direct persuasive style.}
    \label{fig:goodAdExample}
\end{figure}

\begin{figure}[t]
    \centering
    \setlength{\fboxrule}{1pt}
    \fbox{\includegraphics[width=0.8\columnwidth]{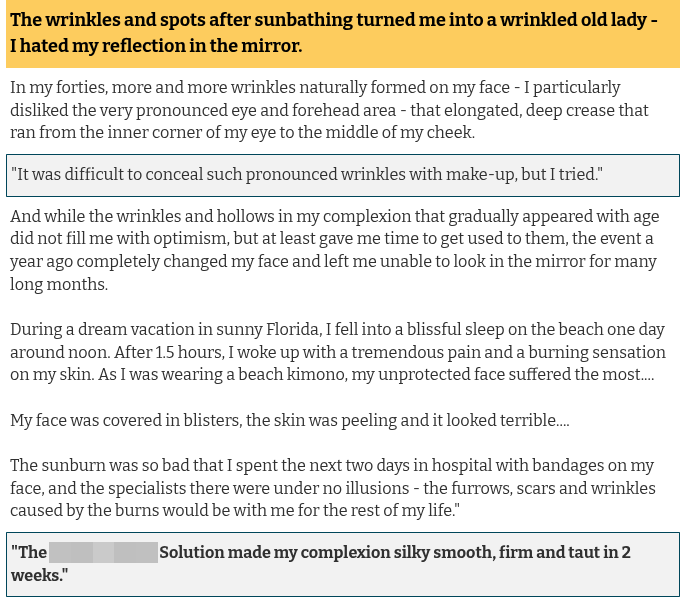}}
    \caption{Example of an advertorial for an anti-aging serum, written in a conversational, informative style and narrating a personal story to promote the product.}
    \Description{Example of an advertorial promoting an anti-aging serum, presented in a conversational and informative tone. The text is structured as a personal narrative, describing an individual's experience and how the serum is used, gradually leading to product promotion in a story-like format.}
    \label{fig:realAdvertorialExample}
\end{figure}

Advertorials differ from traditional advertisements in both tone and style.
While regular ads are direct and promotional, advertorials utilize a more informative or editorial voice, often resembling news articles or feature stories.
This approach engages the reader by providing useful content or storytelling, subtly weaving in the product or service rather than overtly selling it.
We demonstrate the distinctive difference of advertorials in Figures~\ref{fig:goodAdExample} and~\ref{fig:realAdvertorialExample} where two similar products (an anti-wrinkle cream and an anti-aging serum) are promoted using different techniques.
In Figure~\ref{fig:goodAdExample} the traditional ad takes a direct persuasive approach.
It focuses on highlighting the benefits of the cream and urges the audience to buy the product with concise text (\ie ``ADD TO CART'').
On the other hand, the advertorial of Figure~\ref{fig:realAdvertorialExample}, has a more conversational and informative style.
It reads like an article telling the emotional story of a woman who suffered facial damage, only to be saved by the promoted product.
This journalistic tone is a fundamental characteristic of advertorials and was one of the factors that helped annotators reach a decision in the formation of the original dataset, as described in Section~\ref{sec:domain-risk-detection}.
To maintain ethical integrity and academic neutrality, all visible brand names and logos in the advertising screenshots have been blurred.
This prevents unintended promotion or endorsement and avoids potential reputational implications for the depicted companies.

\section{Security Web Engines}
\label{sec:engines-features}

\begin{figure}[t]
    \centering
    \includegraphics[width=\columnwidth]{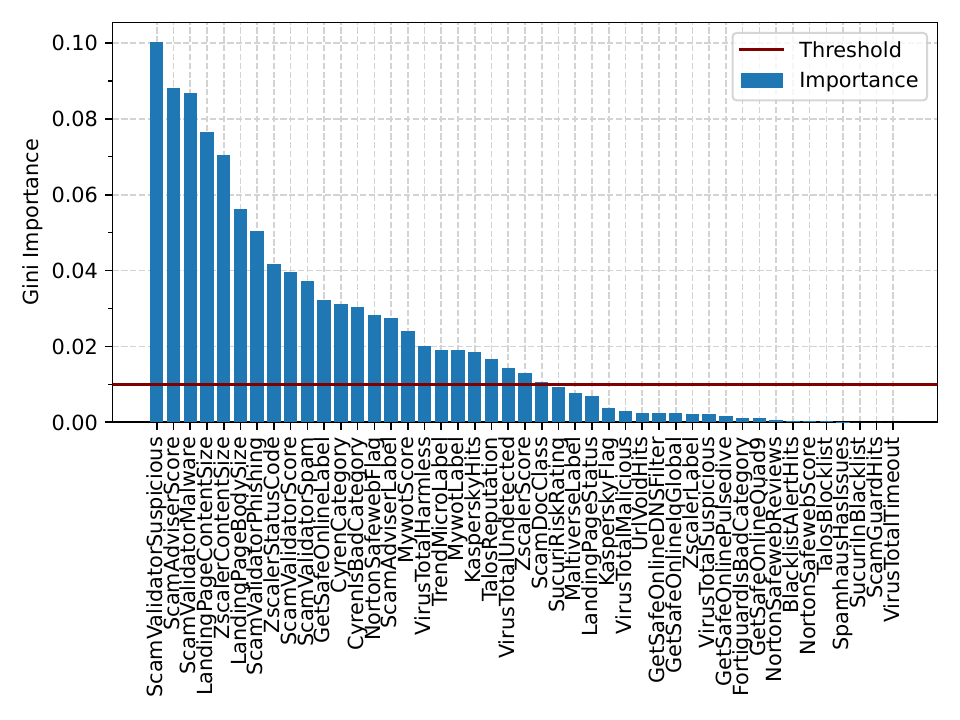}
    \caption{Feature importance of the trained machine learning model for misleading website detection. Feature importance is measured via Gini impurity with 24 features being retained, capturing 96\% of the total importance.}
    \Description{Bar chart showing feature importance scores for detecting misleading websites. Importance is measured using Gini impurity. The 24 retained features account for 96\% of the total importance. The plot ranks features by contribution, highlighting the most influential predictors in the model.}
    \label{fig:featureImportance}
\end{figure}

In order to discover websites that can mislead or deceive users, we extract information from 19 independent security, safety and analytics Web engines (Section~\ref{sec:domain-risk-detection}).
The engines utilized in this work are (in alphabetic order):
(1) Blacklist Alert,
(2) Cyren,
(3) Fortiguard,
(4) GetSafeOnline,
(5) Kaspersky,
(6) Maltiverse,
(7) MyWOT,
(8) Norton SafeWeb,
(9) ScamAdviser,
(10) Scamdoc,
(11) ScamGuard,
(12) Scam Validator,
(13) Spamhaus,
(14) Sucuri,
(15) Talos Intelligence,
(16) Trend Micro,
(17) URLVoid,
(18) VirusTotal,
and (19) Zscaler.
Each web engine provides raw data relevant to the target domain, which is then preprocessed to extract meaningful features.
The raw data from these sources are standardized and transformed through a series of steps including cleaning, normalization and encoding to generate a comprehensive feature set.
To better understand the decision-making process of our model, we compute the feature importance values using Gini impurity and a random forest classifier.
Figure~\ref{fig:featureImportance} presents the ranking of features according to their contribution to the model's classification performance.
We observe that features coming from engines focusing on scam detection (\ie Scam Validator, ScamAdviser) are the most important when detecting misleading websites.
Additionally, the content size of the landing page is a significant feature since it can be associated with malicious intent (Section~\ref{sec:advertorial-content}).
The feature selection process retains only features with meaningful contribution to classification, aiming to reduce noise and redundancy while preserving meaningful information.
The final training dataset consists of 24 features derived from the Web engines.

\section{Examples of Dark Patterns}
\label{sec:dark-patterns-examples}

\begin{figure}[t]
    \centering
    \includegraphics[width=0.8\columnwidth]{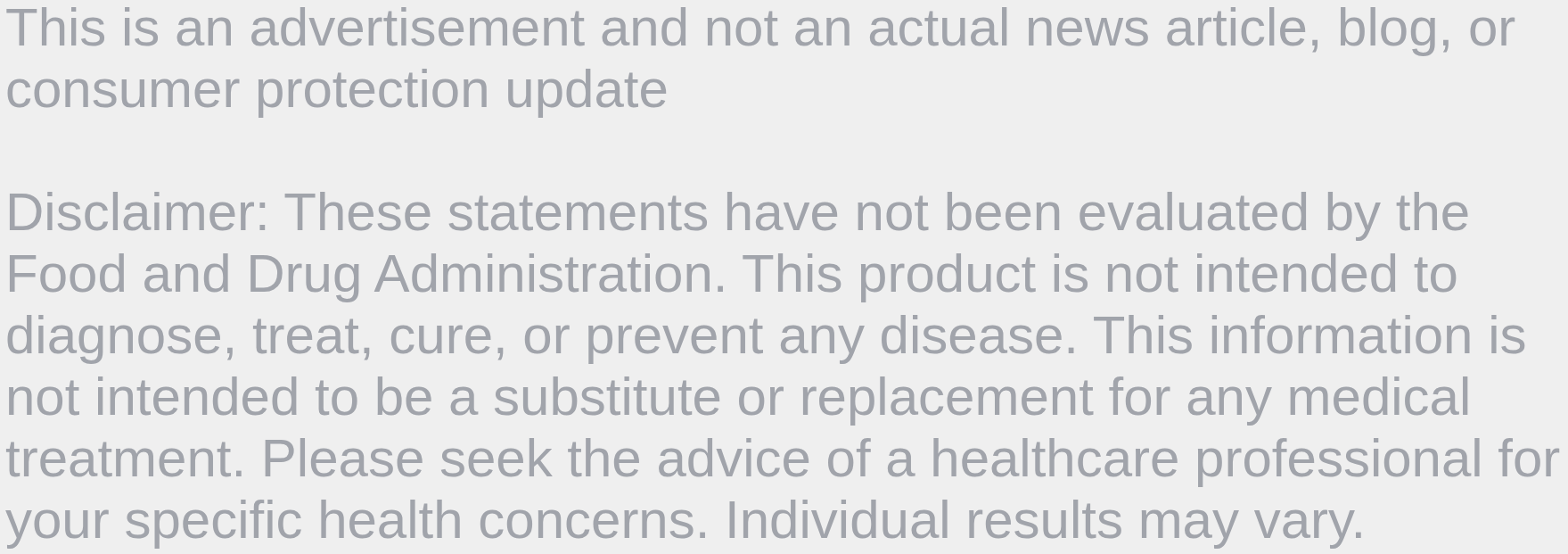}
    \caption{Example of a low-contrast disclaimer text in an online advertorial discovered in this study.}
    \Description{Example of a disclaimer embedded within an online advertorial, displayed in low-contrast text. The disclaimer is less noticeable to readers.}
    \label{fig:lowContrastRatio}
\end{figure}

\begin{figure}[t]
    \centering
    \includegraphics[width=0.8\columnwidth]{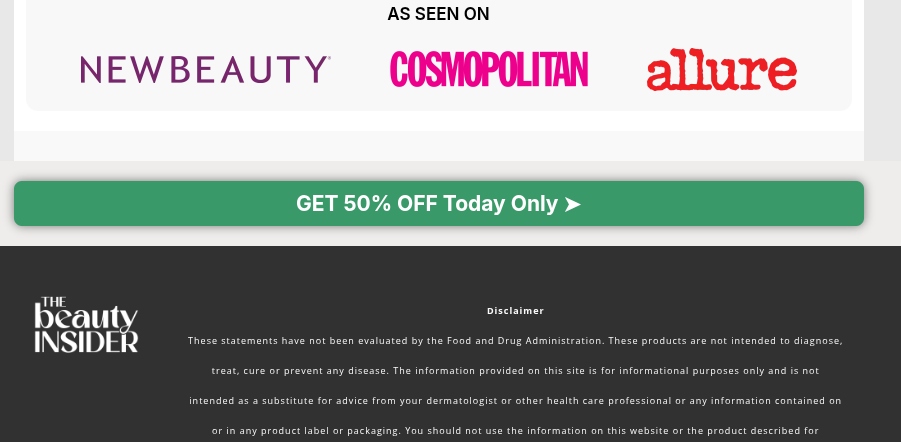}
    \caption{Example of disclaimer in an online advertorial. The disclaimer appears in a font size significantly smaller than the surrounding editorial text.}
    \Description{Example of a disclaimer within an online advertorial, displayed in a font size significantly smaller than the surrounding editorial content.}
    \label{fig:smallFontSize}
\end{figure}

Users often rely on visual and textual signals in order to differentiate between editorial and commercial content and that's why regulation explicitly requires that disclaimers are readily identifiable to indicate the promotional nature of content.
Figure~\ref{fig:lowContrastRatio} provides a real-world example of the \emph{Low-Contrast} dark pattern.
In this example, the disclaimer text is displayed in a color that closely matches the background, making it harder to detect during casual browsing.
The text and background are both in slightly different shades of gray, increasing the possibility that users will overlook the fact they are reading an advertisement and not editorial content.
Next, Figure~\ref{fig:smallFontSize} illustrates how advertorials often present the disclaimer text in a significantly smaller font than the surrounding editorial text.
Users scanning the page may miss the label entirely due to its reduced size and placement at the bottom of the screen, practically turning disclaimers into the ``fine print''.
Both patterns can be regarded as dark patterns because they deliberately reduce the visibility of legally important information (\ie disclaimers).